\newcommand\msout{\bgroup\markoverwith{\textcolor{red}{\rule[0.5ex]{1pt}{1pt}}}\ULon} 

\documentclass[preprint,amsmath,amssymb,aps,showkeys]{revtex4-2}
\usepackage{graphicx}
\usepackage{dcolumn}
\usepackage{bm}
\usepackage{rotating}
\usepackage{multirow,longtable}
\usepackage{ragged2e}
\usepackage{amsthm}
\usepackage{mathtools}
\usepackage{xcolor}
\usepackage{ulem}

\usepackage{pdfpages}

\makeatletter
\AtBeginDocument{\let\LS@rot\@undefined}
\makeatother

\begin{document}


\title{The transition to speciation in the finite genome Derrida-Higgs model: a heuristic solution}


\author{Vitor M. Marquioni$^{1}$}
\email{vitor.MARQUIONI-MONTEIRO@univ-amu.fr}
\author{Marcus A. M. de Aguiar$^1$}
\email{aguiar@ifi.unicamp.br}
\affiliation{$^1$Institute of Physics Gleb Wataghin, University of Campinas, Campinas, SP, Brazil}

\date{\today}

\begin{abstract}

The process of speciation, where an ancestral species divides in two or more new species, involves several geographic, environmental and genetic components that interact in a complex way. Understanding all these elements at once is challenging and simple models can help unveiling the role of each factor separately. The Derrida-Higgs model describes the evolution of a sexually reproducing population subjected to mutations in a well mixed population. Individuals are characterized by a string with entries $\pm 1$ representing a haploid genome with biallelic genes. If mating is restricted by genetic similarity, so that only individuals that are sufficiently similar can mate, sympatric speciation, i.e. the emergence of species without geographic isolation, can occur. Only four parameters rule the dynamics: population size $N$, mutation rate $\mu$, minimum similarity for mating $q_{min}$ and genome size $B$. In the limit $B \rightarrow \infty$, speciation occurs if the simple condition $q_{min} > (1+4\mu N)^{-1}$ is satisfied. However, this condition fails for finite genomes, and speciation does not occur if the genome size is too small. This indicates the existence of a critical genome size for speciation. In this work, we develop an analytical theory of the distribution of similarities between individuals, a quantity that defines how tight or spread out is the genetic content of the population. This theory is carried out in the absence of mating restrictions, where evolution equations for the mean and variance of the similarity distribution can be derived. We then propose a heuristic description of the speciation transition which allows us to numerically calculate the critical genome size for speciation as a function of the other model parameters. The result is in good agreement with the simulations of the model and may guide further investigations on theoretical conditions for species formation.

\end{abstract}

\maketitle


\section{\label{sec:introduction}Introduction}

The emergence of species in the absence of geographic barriers to gene flow is still a very contentious mode of species formation \cite{coyne2004speciation}.
This process, known as sympatric speciation, was proposed by Charles Darwin in his original work of 1859 \cite{turelli2001theory} but remained an unsolved problem in evolutionary biology \cite{coyne2004speciation,bolnick2007sympatric,gavrilets2003perspective,martin2022sympatric}, with even its formal definition being a matter of debate \cite{gavrilets2003perspective,fitzpatrick2008if,mallet2009space}.
As a consequence, it is still unclear under which conditions different species can evolve from an ancestor in the presence of constant gene flow. It is also quite difficult to verify if two species have actually emerged via sympatric speciation or not \cite{martin2022sympatric,sun2022sympatric,kautt2016multispecies}.
In 2004, Coyne and Orr \cite{coyne2004speciation} established four criteria to characterize this process in nature, but with the advances in molecular biology, it has become even more challenging to certify evolution in sympatry and questions over previously known examples appeared \cite{fitzpatrick2008if,martin2015complex,richards2018don,sun2022sympatric}.
Many cichlid species seem to have emerged in sympatry in different glacial lakes, such as in the Great African lakes \cite{schliewen1994sympatric,seehausen1999can} and in Central American lakes \cite{barluenga2004midas,barluenga2006sympatric,elmer2010local}. 
These fishes constitute a group of species that has passed through the four Coyne and Orr's criteria, but even this standard example casts some doubts when the genomic data is deeply analyzed \cite{martin2015complex}.

Mathematical and computational models have tried to unravel the conditions behind sympatric speciation, and selection seems to play an important role in this process \cite{dieckmann1999origin,gavrilets2006maynard,bolnick2007sympatric}. However, in 1991, the Derrida-Higgs model \cite{higgs1991stochastic} shed some light on how a \textit{neutral} population evolving in sympatry could give rise to reproductively isolated groups, hence called species. According to this model, sexual reproduction and genetic drift counterbalance the effects of mutations, resulting in a genetically similar population that does not split into species. Nevertheless, when mating is restricted by genetic similarity, so that individuals that are too dissimilar cannot mate, gene flow is reduced and groups of individuals that cannot crossbreed can appear. In the limit of infinitely many genes, the authors conjectured a sharp transition between the regimes of single and multiple species, which is well observed in simulations \cite{higgs1991stochastic,higgs1992genetic}. In 2017, de Aguiar \cite{de2017speciation} numerically showed that this transition is not satisfied for finite genomes, indicating that the number of genes poses a non-trivial barrier to sympatric speciation.

For small numbers of genes, the average similarity between individuals converges to the reproduction similarity threshold and does not evolve beyond this limit \cite{de2017speciation}, hindering speciation. In this case, there is gene flow between almost every pair of individuals in the community, defining a single species regime. However, for large enough genome sizes, the average similarity among the individuals evolves to values below the similarity threshold and isolated groups, i.e. groups of individuals with no gene flow between them, appear, approaching the situation with infinitely many genes. Spatial versions of the Derrida-Higgs model \cite{de2009global} also show dependencies on the genome size \cite{de2017speciation}, displaying important effects on the structure of the underlying phylogenetic trees \cite{costa2019signatures}.

In spite of the evident role of the genome size on the transition from one to multiple species in the Derrida-Higgs model, no analytical theory to describe such dependency has been developed so far. In particular, the value of the smallest genome size that allows for a many species regime is not known. To formulate such theory is the aim of this work. We introduce exact equations for the evolution of the genetic similarity distribution of the population, which we solve for the first and second moments in the absence of mating restrictions. We then propose a heuristic explanation to the speciation transition and use the previously obtained results to calculate the critical number of genes that allows for species formation. 

The remainder of this paper is organized as: in section \ref{sec:model} we review the Derrida-Higgs model and we introduce the analytical theory of the similarity distribution in section \ref{sec:similarity}. The analytical calculation of mean and variance is performed in section \ref{sec:mean}, which is followed by section \ref{sec:allresults} containing all the mathematical results. The detailed calculations are left to the supplementary material. In section \ref{sec:heuristic}, the heuristic solution is proposed and the critical genome size is calculated. Finally, simulations comparing the analytical results are presented in section \ref{sec:solution}.


\section{\label{sec:model}The Model - a brief review}

We consider a population of size $N$ whose individuals $\alpha$ are described at time $t$ by their genomes $\mathbf{S}^{\alpha}_t=\{s_{t,1}^{\alpha},\ldots,s_{t,B}^{\alpha}\}$, a binary sequence of length $B$, with $s_{t,i}^{\alpha}=\pm1$. The similarity between two individuals $\alpha$ and $\beta$ is defined by
\begin{equation}
    q_t^{\alpha\beta}=\frac{1}{B}\sum_{i=1}^Bs_{t,i}^{\alpha}s_{t,i}^{\beta},
    \label{eqsim}
\end{equation}
and it measures how genetically similar a pair of individuals is. If $q_t^{\alpha\beta}=1$, then $\alpha$ and $\beta$ are identical, whereas $q_t^{\alpha\beta}=0$ means that half of their genes is different. The similarity can also be written in terms of the genetic distance $d^{\alpha\beta}_t$, which counts how many alleles differ between $\alpha$ and $\beta$, i.e., the Hamming distance between the sequences $\mathbf{S}^{\alpha}_t$ and $\mathbf{S}^{\beta}_t$:
\begin{equation}
    d^{\alpha\beta}_t=\frac{1}{2}\sum_{i=1}^B|s_{t,i}^{\alpha}-s_{t,i}^{\beta}|=\frac{B}{2}(1-q_{t}^{\alpha\beta}).
\end{equation}

In each generation $t$, $N$ focal individuals $\alpha$ are chosen at random, with replacement, and a mating partner $\beta$ for each one of them is randomly chosen among the set of individuals satisfying $q_t^{\alpha\beta}\ge q_{min}$. The assortative parameter $q_{min}$ prevents individuals that are not sufficiently similar to reproduce. Every mating event gives rise to an individual $\gamma$ for generation $t+1$, described by a binary sequence which has every allele coming from one parent or from the other, with the same probability. Also, with a mutation rate $\mu$, an allele changes its sign. In the time interval of one generation, the probability that the allele changes sign is $(1-e^{2\mu})/2$. The sexual reproduction tends to homogenize the genetic diversity and because the population size is kept constant, rare mutants are likely to die with no descendants (genetic drift). The mutation rate is then the only source of diversity, whose effects are counterbalanced by drift and reproduction.

Derrida and Higgs \cite{higgs1991stochastic} considered infinitely large genomes ($B\rightarrow\infty$) for which they showed that, in the absence of mating restrictions ($q_{min}\rightarrow-1$), the average of the similarity distribution, i.e. the average of the similarity over all pairs of individuals, should follow
\begin{equation}
    \langle q_{t+1}^{\alpha\beta}\rangle=\frac{e^{-4\mu}}{N}\left(1+(N-1)\langle q_{t}^{\alpha\beta}\rangle\right),\label{eq:mean}
\end{equation}
which has the equilibrium
\begin{equation}
    q_{eq}=\frac{1}{Ne^{4\mu}-(N-1)}\approx\frac{1}{1+4\mu N},
\end{equation}
where the approximation holds for $N \gg 1$ and $\mu\ll1$.

If mating restrictions are imposed, the distribution of similarities still has a tendency to settle at $q_{eq}$. However, if $q_{min} > q_{eq}$, as soon as $\langle q_t^{\alpha\beta}\rangle$ reaches $q_{min}$, there will be pairs of individuals with $q_t^{\alpha\beta} < q_{min}$ that will not be able to reproduce. The consequence of this conflict, between trying to reach $q_{eq}$ while keeping $q_t^{\alpha\beta} > q_{min}$, is the break up of the population into isolated groups: within a group, individuals may reproduce, but there is no gene flow between any pair of individuals that belong to different groups. Thus such groups define different species and the speciation transition happens as long as $q_{min}>q_{eq}$. In a network theory language, at time $t$, every individual represents the node of a network of size $N$ and the nodes are connected if and only if $q_{t}^{\alpha\beta}\ge q_{min}$. Evolution changes the number of connections and creates communities: if $q_{min}>q_{eq}$, given enough time, the network will break up into non-connected components.

However, for finite genome sizes ($B<\infty$), the variance of the distribution of similarities plays a role and, if it is large enough, although many pairs of individuals cannot reproduce, the connected pairs are still able to generate a connected community in the following generation, even if $q_{min}>q_{eq}$. The similarity distribution gets narrower for increasing genome sizes, and for values larger than a critical value $B_c$, the formation of species is observed after a finite number of generations. This behaviour was first studied in 2017 by de Aguiar \cite{de2017speciation} but no analytical theory was proposed.

\section{The similarity distribution}
\label{sec:similarity}

We start by constructing the similarity probability distribution of two individuals $\alpha$ and $\beta$ at generation $t+1$. The parents of $\alpha$ are individuals $p_1$ and $p_2$, and the parents of $\beta$ are $p_1'$ and $p_2'$, and they all belong to generation $t$, (we can now suppress the time index since they are always followed by the index of the individual). Supposing $s_{i}^{\alpha}$ comes from $p_1$, the probability that it is equal to $s_{i}^{p_1}$, i.e., that it does not mutate, can be written as
\begin{equation}
    \mathcal{P}\left(s_i^{\alpha}|p_1\right)=\frac{1}{2}\left[1+(2r^c-1)s_i^{\alpha}s_i^{p_1}\right],
\end{equation}
where $r^c=\frac{1}{2}(1+e^{-2\mu})$ is the probability of no mutation. Therefore, given the pair of parents $(p_1,p_2)$ of $\alpha$ and that the allele $i$ has equal chances of been inherited from $p_1$ and $p_2$,
\begin{equation}
    \mathcal{P}\left(s_i^{\alpha}|(p_1,p_2)\right)=\frac{1}{2}+s_i^{\alpha}\frac{e^{-2\mu}}{4}\left(s_i^{p_1}+s_i^{p_2}\right).
\end{equation}

Because the alleles are independent, for the entire genome of $\alpha$,
\begin{equation}
    \mathcal{P}\left(\mathbf{S}^{\alpha}|(p_1,p_2)\right)=\prod_{i=1}^B\left[\frac{1}{2}+s_i^{\alpha}\frac{e^{-2\mu}}{4}\left(s_i^{p_1}+s_i^{p_2}\right)\right],
\end{equation}
and similarly for $\beta$.

Now, if the genomes of $\alpha$ and $\beta$ are known, their similarity is uniquely defined, thus
\begin{equation}
    \mathcal{P}(q^{\alpha\beta}|\mathbf{S}^{\alpha},\mathbf{S}^{\beta})=\delta\left(q^{\alpha\beta},\mathbf{S}^{\alpha}\cdot\mathbf{S}^{\beta}/B\right),
\end{equation}
where $\delta$ is the Dirac $\delta$ and $\cdot$ is the usual scalar product. For the total probability one may write
\begin{equation}
    \mathcal{P}(q^{\alpha\beta})=\sum_{\mathbf{S}^{\alpha},\mathbf{S}^{\beta}}\mathcal{P}(q^{\alpha\beta}|\mathbf{S}^{\alpha},\mathbf{S}^{\beta})\mathcal{P}(\mathbf{S}^{\alpha},\mathbf{S}^{\beta}),
\end{equation}
with the sum running over all the possible genomes $\mathbf{S}^{\alpha}$ and $\mathbf{S}^{\beta}$. This equation leaves us with the task of calculating the joint probability of the genomes of two individuals $\alpha$ and $\beta$. Conditioning it to the genome of their parents, we can write
\begin{equation}
    \mathcal{P}(\mathbf{S}^{\alpha},\mathbf{S}^{\beta})=\sum_{(p_1,p_2)}\sum_{(p_1',p_2')}\mathcal{P}(\mathbf{S}^{\alpha},\mathbf{S}^{\beta}|(p_1,p_2),(p_1',p_2'))\mathcal{P}((p_1,p_2),(p_1',p_2')),
\end{equation}
where the sums run over all the possible pairs of parents of $\alpha$ and $\beta$,
\begin{equation}
    \mathcal{P}(\mathbf{S}^{\alpha},\mathbf{S}^{\beta}|(p_1,p_2),(p_1',p_2'))=\mathcal{P}(\mathbf{S}^{\alpha}|(p_1,p_2))\mathcal{P}(\mathbf{S}^{\beta}|(p_1',p_2')),
\end{equation}
and
\begin{equation}
    \mathcal{P}((p_1,p_2),(p_1',p_2'))=\mathcal{P}((p_1,p_2))\mathcal{P}((p_1',p_2')),
\end{equation}
because the parents of $\alpha$ and $\beta$ are independently drawn.

The last task is to calculate the probability $\mathcal{P}((p_1,p_2))$ of a given pair to be drawn from the population at generation $t$. Notice that the notation does not distinguish between which individual is the focal and which individual is drawn afterwards to be its mate. Since the focal is chosen randomly, the probability of a given individual to be focal is given by $1/N$. Let us define the adjacency matrix $\mathbb{A}$ of the population at time $t$, whose elements are $A_{ij}=1$, if $q_t^{ij}\ge q_{min}$ for $i\ne j$, and 0 otherwise. The mating partner of the focal individual is randomly chosen from the set of individuals who are connected to the focal. Calling the degree of a node $i$ as $N_i=\sum_{j}A_{ij}$, we find
\begin{equation}
    \mathcal{P}((p_1,p_2))=\frac{A_{p_1p_2}}{N}\left(\frac{1}{N_{p_1}}+\frac{1}{N_{p_2}}\right),
\end{equation}
whose details can be found in the supplementary material. Figure \ref{fig:networks} shows how the network defined by the community changes throughout the dynamics and the non-connected components appear.

Combining these results and rearranging the sums, we find the probability distribution for the similarity $q^{\alpha\beta}$ written in terms as the genomes of the individuals from the previous generation,
\begin{align}
    \mathcal{P}(q_{t+1}^{\alpha\beta})=&\frac{1}{N^2}\sum_{\mathbf{S}^{\alpha},\mathbf{S}^{\beta}}\sum_{p_1,p_2}\sum_{p_1',p_2'}\delta\left(q_{t+1}^{\alpha\beta},\mathbf{S}^{\alpha}\cdot\mathbf{S}^{\beta}/B\right)\frac{A_{p_1p_2}A_{p_1'p_2'}}{N_{p_1}N_{p_1'}}\times\nonumber\\
    &\times\prod_{i=1}^B\left[\frac{1}{2}+\frac{s_{i,t+1}^{\alpha}e^{-2\mu}}{4}\left(s_{i,t}^{p_1}+s_{i,t}^{p_2}\right)\right]
    \left[\frac{1}{2}+\frac{s_{i,t+1}^{\beta}e^{-2\mu}}{4}\left(s_{i,t}^{p_1'}+s_{i,t}^{p_2'}\right)\right],
    \label{eqgen}
\end{align}
in which the sums over $p_1$, $p_2$, $p_1'$ and $p_2'$ run on all the individuals of the generation $t$. From this result, we are able to calculate the evolution of the average similarity of the population even when $B<\infty$.

\begin{figure}
    \centering
    \includegraphics[width=0.9\linewidth]{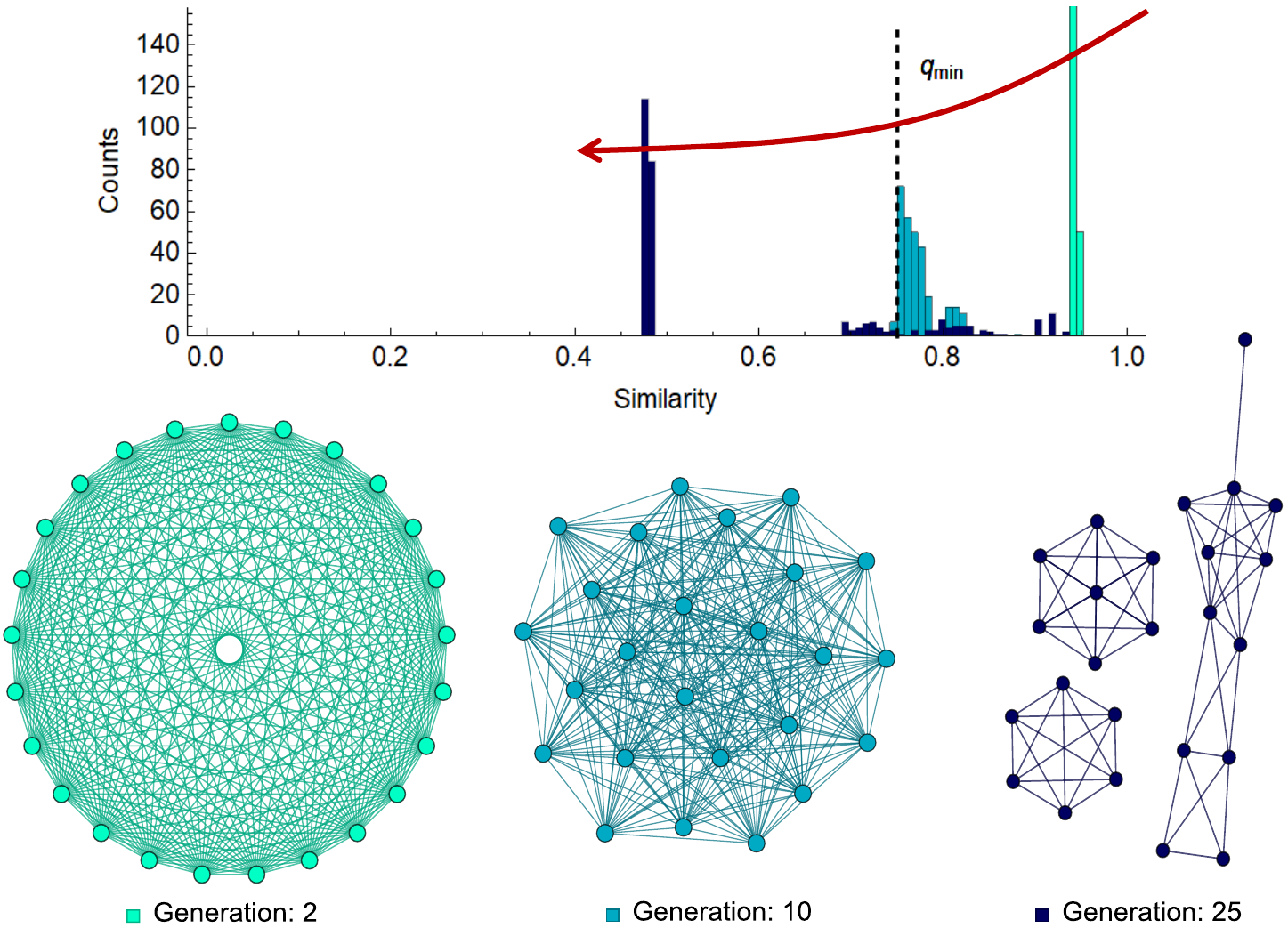}
    \caption{\textbf{Network visualization of the Derrida-Higgs dynamics.} Evolution of a population of $N=25$ individuals with mutation rate $\mu=0.008$ under the Derrida-Higgs dynamics. The plot shows the similarity distribution in 3 different generations starting from a clonal population. The distribution moves toward smaller values of similarity breaking into different peaks after crossing the reproduction threshold $q_{min}=0.75$. The underlying network is represented at each generation, going from a complete network to a network with three components.}
    \label{fig:networks}
\end{figure}

\section{Mean and Variance\label{sec:mean}}

In this section, we use Eq.(\ref{eqgen}) to calculate the time evolution of the average similarity and the variance. Other analytical results are presented in Section \ref{sec:allresults} and all detailed calculations are in the supplementary material.

\begin{figure}[t]
    \centering
    \includegraphics[width=1\linewidth]{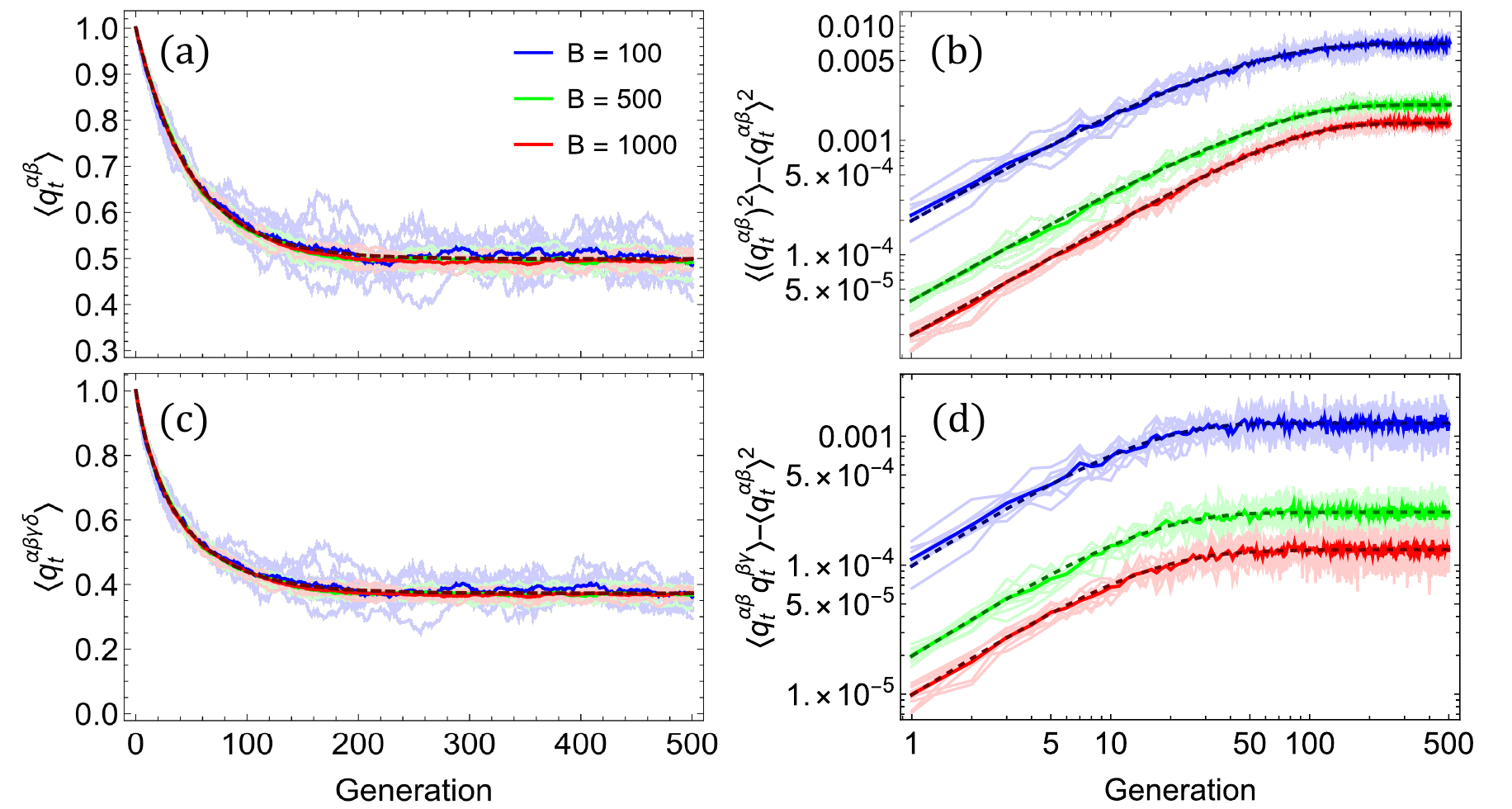}
    \caption{\textbf{Evolution of the first and second moments of the similarity distribution.} In the figure, the average (dark curves) over 10 simulations (light curves) of the first and second moments of the similarity distribution are presented and contrasted with the analytical results (dashed curves). The simulations consider no mating restrictions. Panel (a) shows the mean similarity (Eq.\eqref{eq:mean}), panel (b) shows the variance (Eq.\eqref{eq:variancia}); panel (c) shows the evolution of the mean second order overlap (Eq.\eqref{eq:secondoverlap}) and panel (d) shows the covariance of the similarity distribution (Eq.\eqref{eq:covariancia}). The average second order overlap was calculated from the simulations by considering a random sample of size $N^2$ from the set of all second order overlaps (whose size is $\sim N^4$). The parameters of the simulations are $N=100$ and $\mu=0.0025$, and the genome sizes are indicated in the figure.}
    \label{fig:analytics}
\end{figure}

Given the similarity values at a time $t$, the expected similarity at the next generation is given by
\begin{align}
    \mathbb{E}(q_{t+1}^{\alpha\beta})=\frac{e^{-4\mu}}{4N^2}\sum_{p_1,p_2}\sum_{p'_1,p'_2}\frac{A_{p_1p_2}A_{p'_1p'_2}}{N_{p_1}N_{p'_1}}\left(q_t^{p_1p_1'}+q_t^{p_1p_2'}+q_t^{p_2p_1'}+q_t^{p_2p_2'}\right),	\label{eq:expected}
\end{align}
where use used the delta function to kill the sums over $\mathbf{S}^{\alpha}$ and $\mathbf{S}^{\beta}$ and also the definition
of similarity, Eq.(\ref{eqsim}). When there are no restrictions to mating, $A_{ij}=1$ for $i\ne j$ and $N_i=N-1$, the result simplifies to:
\begin{align}
    \langle q_{t+1}^{\alpha\beta}\rangle=e^{-4\mu}\left[\frac{1}{N}+\left(1-\frac{1}{N}\right)\langle q_t^{\alpha\beta}\rangle\right],\label{eq:mean}
\end{align}
where $\langle\cdot\rangle$ is the ensemble average. Notice that for the average there is no explicit dependence on the genome size, and thus the equilibrium of average similarity when there are no restrictions to mating is the same regardless of the genome size. The panel (a) of Figure \ref{fig:analytics} shows the average of many numerical simulations in comparison with the prediction of this equation. Although the deviations from the average are larger for smaller genome sizes, the average mean similarity does not depend on $B$.

The second moment of the distribution, however, does depend explicitly on the genome size:
\begin{align}
    \mathbb{E}((q_{t+1}^{\alpha\beta})^2) &=\frac{1}{N^2}\sum_{p_1,p_2}\sum_{p'_1,p'_2}\frac{A_{p_1p_2}A_{p'_1p'_2}}{N_{p_1}N_{p'_1}}\left[\frac{1}{B}+\frac{e^{-8\mu}}{16}\left[(q_t^{p_1p_1'}+q_t^{p_1p_2'}+q_t^{p_2p_1'}+q_t^{p_2p_2'})\right]^2\right.\nonumber\\
    &\left.-\frac{e^{-8\mu}}{4B}\left(1+q_t^{p_1p_2}+q_t^{p_1'p_2'}+q_t^{p_1p_2p_1'p_2'}\right)\right],		
\end{align}
where the \textit{second order overlap} is defined by
\begin{equation}
    q^{\alpha\beta\gamma\delta}=\frac{1}{B}\sum_{i=1}^Bs_i^{\alpha}s_i^{\beta}s_i^{\gamma}s_i^{\delta}.
\end{equation}

Although this equation is very cumbersome to analyse, in the absence of mating restrictions the variance at any moment in time can be written as
\begin{equation}
    \sigma_{B,t}^2=\left\langle \mathbb{E}((q_{t}^{\alpha\beta})^2)- \mathbb{E}(q_{t}^{\alpha\beta})^2\right\rangle=\Lambda_1(t)+\frac{\Lambda_2(t)}{B},\label{eq:sigma}
\end{equation}
where $\Lambda_1$ and $\Lambda_2$ do not depend on  $B$ (see Supl. Material). Therefore, the effects of the genome size on the similarity distribution diminishes as the number of alleles increases, with the distribution being narrower for larger genomes. The panel (b) of Figure \ref{fig:analytics} shows the computational results and the analytical solutions reported in this paper for the variance.


\newpage

\section{A summary of the analytical results \label{sec:allresults}}

This section summarizes all analytical results calculated with the present theory and their detailed calculations can be found in the supplementary material.

	Given a population of $N$ individuals, for every $\alpha\ne\beta$, if $q_t^{\alpha\beta}\ge q_{min}$, $A_{\alpha\beta}=1$ and 0 otherwise; $N_{\alpha}=\sum_{\beta}A_{\alpha\beta}$, i.e., the matrix $\mathbb{A}$ with elements $(\mathbb{A})_{\alpha\beta}=A_{\alpha\beta}$ is the adjacency matrix of the underlying network and $N_{\alpha}$ is the degree of the vertice $\alpha$. The individuals $\alpha$, $\beta$, $\gamma$ and $\delta$ are different individuals of generation $t+1$, whose parents are respectively $(p_1,p_2)$, $(p_1',p_2')$, $(p_1'',p_2'')$ and $(p_1''',p_2''')$. The genome size is $B$ and the mutation rate $\mu$.

	\begin{enumerate}
		\item \textbf{Similarity Distribution}. The probability distribution of a similarity $q_{t+1}^{\alpha\beta}$ is given by
		\begin{align}
			\mathcal{P}(q_{t+1}^{\alpha\beta})=&\frac{1}{N^2}\sum_{\mathbf{S}^{\alpha},\mathbf{S}^{\beta}}\sum_{p_1,p_2}\sum_{p'_1,p'_2}\delta\left(q_{t+1}^{\alpha\beta},\frac{\mathbf{S}^{\alpha}\cdot\mathbf{S}^{\beta}}{B}\right)\frac{A_{p_1p_2}A_{p'_1p'_2}}{N_{p_1}N_{p_1'}} \times\nonumber\\\times&\prod_{i=1}^B\left[\frac{1}{2}+\frac{s_{i,t+1}^{\alpha} e^{-2\mu}}{4}\left(s_{i,t}^{p_1}+s_{i,t}^{p_2}\right)\right]\left[\frac{1}{2}+\frac{s_{i,t+1}^{\beta} e^{-2\mu}}{4}\left(s_{i,t}^{p_1'}+s_{i,t}^{p_2'}\right)\right].
		\end{align}
		\item \textbf{Expected Similarity}. Given the similarity values at a time $t$, the expected similarity at the next generation is given by
		\begin{align}
			\mathbb{E}(q_{t+1}^{\alpha\beta})=\frac{e^{-4\mu}}{4N^2}\sum_{p_1,p_2}\sum_{p'_1,p'_2}\frac{A_{p_1p_2}A_{p'_1p'_2}}{N_{p_1}N_{p'_1}}\left(q_t^{p_1p_1'}+q_t^{p_1p_2'}+q_t^{p_2p_1'}+q_t^{p_2p_2'}\right),	
		\end{align}
		and when there are no restrictions to mating
		\begin{align}
			\langle q_{t+1}^{\alpha\beta}\rangle=e^{-4\mu}\left[\frac{1}{N}+\left(1-\frac{1}{N}\right)\langle q_t^{\alpha\beta}\rangle\right].\label{eq:meanAppendix}
		\end{align}
		
		\item  \textbf{The second moment of the similarity distribution.} The expected value of the second moment of the similarity distribution $\mathcal{P}(q_{t+1}^{\alpha\beta})$ is given by
		\begin{align}
			\mathbb{E}((q_{t+1}^{\alpha\beta})^2) &=\frac{1}{N^2}\sum_{p_1,p_2}\sum_{p'_1,p'_2}\frac{A_{p_1p_2}A_{p'_1p'_2}}{N_{p_1}N_{p'_1}}\left[\frac{1}{B}+\frac{e^{-8\mu}}{16}\left[(q_t^{p_1p_1'}+q_t^{p_1p_2'}+q_t^{p_2p_1'}+q_t^{p_2p_2'})\right]^2\right.\nonumber\\
			&\left.-\frac{e^{-8\mu}}{4B}\left(1+q_t^{p_1p_2}+q_t^{p_1'p_2'}+q_t^{p_1p_2p_1'p_2'}\right)\right],		\end{align}
		where the \textit{second order overlap} is defined by
		\begin{equation}
			q^{\alpha\beta\gamma\delta}=\frac{1}{B}\sum_{i=1}^Bs_i^{\alpha}s_i^{\beta}s_i^{\gamma}s_i^{\delta}.
		\end{equation}
		
		\item \textbf{The second order overlap.} Once we have defined the second order overlap, we can calculate its expected value,
		\begin{align}
			\mathbb{E}(q_{t+1}^{\alpha\beta\gamma\delta})&=\frac{e^{-8\mu}}{16N^4}\sum_{p_1,p_2}\sum_{p_1',p_2'}\sum_{p_1'',p_2''}\sum_{p_1''',p_2'''}\frac{A_{p_1p_2}A_{p_1'p_2'}A_{p_1''p_2''}A_{p_1'''p_2'''}}{N_{p_1}N_{p_1'}N_{p_1''}N_{p_1'''}}\nonumber\\
			&\times
			\left(q_t^{p_1p_1'p_1''p_1'''}+q_t^{p_1p_1'p_1''p_2'''}+q_t^{p_1p_1'p_2''p_1'''}+q_t^{p_1p_1'p_2''p_2'''}+q_t^{p_1p_2'p_1''p_1'''}+q_t^{p_1p_2'p_1''p_2'''}\right.\nonumber\\
			&+q_t^{p_1p_2'p_2''p_1'''}+q_t^{p_1p_2'p_2''p_2'''}+q_t^{p_2p_1'p_1''p_1'''}+q_t^{p_2p_1'p_1''p_2'''}+q_t^{p_2p_1'p_2''p_1'''}\nonumber\\
			&\left.+q_t^{p_2p_1'p_2''p_2'''}+q_t^{p_2p_2'p_1''p_1'''}+q_t^{p_2p_2'p_1''p_2'''}+q_t^{p_2p_1'p_2''p_2'''}+q_t^{p_2p_2'p_2''p_2'''}\right),
		\end{align}
		and in the absence of restrictions to mating,
		\begin{align}
			\langle q_{t+1}^{\alpha\beta\gamma\delta}\rangle&=\frac{e^{-8\mu}}{N^3}\left[(3N-2)+(N-1)(6N-8)\langle q_t^{\alpha\beta}\rangle+(N-1)(N-2)(N-3)\langle q_t^{\alpha\beta\gamma\delta}\rangle\right].\label{eq:secondoverlap}
		\end{align}
            The panel (c) of Figure \ref{fig:analytics} compares the analytical description and computational results for the second order overlap.
  
		\item  \textbf{The variance evolution.} If there are no mating restrictions, the variance of the similarity distribution evolves according to
		\begin{align}
			&\textrm{Var}(q_{t+1}^{\alpha\beta})\nonumber\\
			&=\frac{1}{B}-\frac{e^{-8\mu}}{4B}\left[\left(1+\frac{2}{N(N-1)}\right)+\left(2+\frac{4(N-2)}{N(N-1)}\right)\langle q_t^{\alpha\beta}\rangle+\frac{(N-2)(N-3)}{N(N-1)}\langle q_t^{\alpha\beta\gamma\delta}\rangle\right]\nonumber\\
			&+\frac{e^{-8\mu}(N-2)^2}{4N^2(N-1)}\left[\left(1-\langle q_t^{\alpha\beta}\rangle\right)^2+\left(N+2-\frac{2}{N}\right)\textrm{Var}(q_t^{\alpha\beta})\right.+\left.2\left(N-6+\frac{4}{N}\right)\textrm{Cov}(t)^{\alpha\beta\gamma}\right],\label{eq:variancia}
		\end{align}
		in which we have also defined the covariance
		\begin{align}
			\textrm{Cov}(t)^{\alpha\beta\gamma}&=\textrm{Cov}(q_t^{\alpha\beta},q_t^{\beta\gamma}).
		\end{align}
		
		\item  \textbf{The covariance evolution with a common individual.} The covariance between two similarities with one individual in common, $\textrm{Cov}(t)^{\alpha\beta\gamma}$, in the absence of mating restrictions, evolves as
		\begin{align}
			\textrm{Cov}(t+1)^{\alpha\beta\gamma}
			&=\frac{e^{-4\mu}}{B}\left[\frac{1}{N}+\left(1-\frac{1}{N}\right)\langle q_t^{\alpha\beta}\rangle\right]\nonumber\\
			&-\frac{e^{-8\mu}}{2B}\left[\frac{2}{N^2}+\frac{1}{N}+\left(1+\frac{4}{N}-\frac{8}{N^2}\right)\langle q_t^{\alpha\beta}\rangle+\left(1-\frac{2}{N}\right)\left(1-\frac{3}{N}\right)\langle q_t^{\alpha\beta\gamma\delta}\rangle\right]\nonumber\\
			&+\frac{e^{-8\mu}(N-2)^2}{2N^3}\left[\textrm{Var}(q_t^{\alpha\beta})+(N-4)\textrm{Cov}(t)^{\alpha\beta\gamma}\right].\label{eq:covariancia}
		\end{align}
            Comparisons of this equation and computational simulations are presented in the panel (d) of Fig. \ref{fig:analytics}.

		\item  \textbf{The covariance with no common individual.} The covariance between similarities that do not share any common individual is zero,
		\begin{equation}
			\textrm{Cov}(t)^{\alpha\beta\gamma\delta}=	\mathbb{E}(q_{t+1}^{\alpha\beta}q_{t+1}^{\gamma\delta})-\mathbb{E}(q_{t+1}^{\alpha\beta})\mathbb{E}(q_{t+1}^{\gamma\delta})=0.
		\end{equation}
	\end{enumerate}

Mean and Variance in the limit $B\rightarrow\infty$ and $N\gg1$:

        \begin{enumerate}
		\item \textbf{Expected Similarity}. In this limit, the result does not change,
		\begin{align}
			\langle q_{t+1}^{\alpha\beta}\rangle=e^{-4\mu}\left[\frac{1}{N}+\left(1-\frac{1}{N}\right)\langle q_t^{\alpha\beta}\rangle\right].
			\nonumber
		\end{align}
		
		\item \textbf{The second order overlap.}
		\begin{align}
			\langle q_{t+1}^{\alpha\beta\gamma\delta}\rangle&= e^{-8\mu}\left[\frac{6}{N}\langle q_t^{\alpha\beta}\rangle+\left(1-\frac{6}{N}\right)\langle q_t^{\alpha\beta\gamma\delta}\rangle\right].
		\end{align}
		
		\item  \textbf{The variance evolution.}
		\begin{align}
			&\textrm{Var}(q_{t+1}^{\alpha\beta})=\frac{e^{-8\mu}}{4}\left[\frac{1}{N}\left(1-\langle q_t^{\alpha\beta}\rangle\right)^2+\left(1-\frac{1}{N}\right)\textrm{Var}(q_t^{\alpha\beta})+2\left(1-\frac{9}{N}\right)\textrm{Cov}(t)^{\alpha\beta\gamma}\right].
		\end{align}
		
		\item  \textbf{The covariance evolution with a common individual.}
		\begin{align}
			\textrm{Cov}(t+1)^{\alpha\beta\gamma}
			&=\frac{e^{-8\mu}}{2}\left[\frac{1}{N}\textrm{Var}(q_t^{\alpha\beta})+\left(1-\frac{8}{N}\right)\textrm{Cov}(t)^{\alpha\beta\gamma}\right].
		\end{align}
	\end{enumerate}

\section{Heuristic description of the speciation transition \label{sec:heuristic}}

The theory developed in the previous sections is exact and can be used to calculate the moments of the similarity distribution when mating is not restricted. In this case, however, there is no species formation. When the mating restriction $q_{min}$ is introduced, the theory can still be used to describe the evolution of the population if the similarity between the individuals is much larger than the reproduction threshold ($\langle q^{\alpha\beta}\rangle>q_{min}$). Hence, we use such equations to describe the system up to the speciation transition, when $\langle q^{\alpha\beta}\rangle \approx q_{min}$.

\begin{figure}[tb]
    \centering
    \includegraphics[width=1\linewidth]{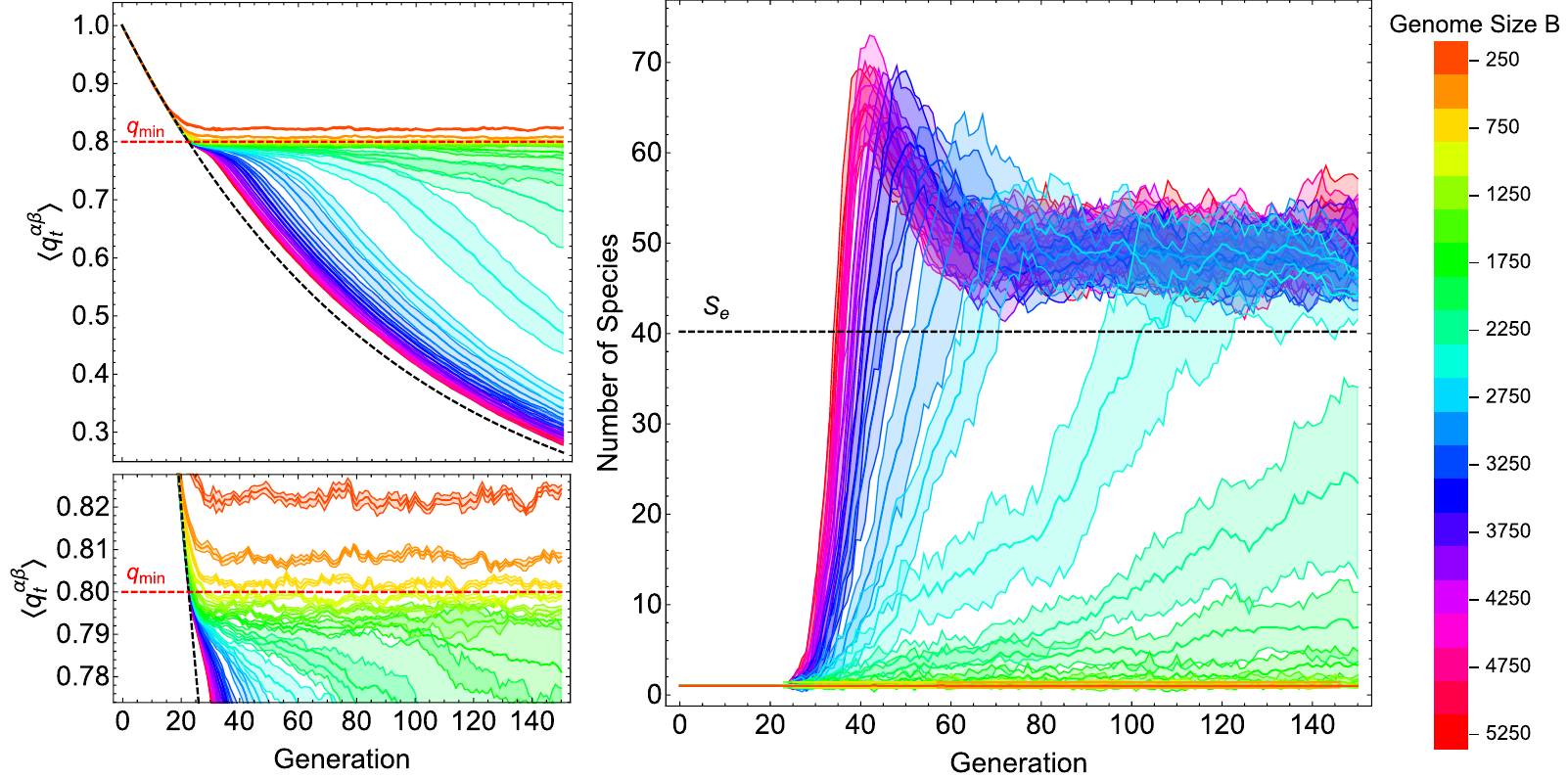}
    \caption{\textbf{Species formation and the genome size.} The top left panel shows the evolution of the average similarity for different genome sizes $B$. Each curve is the average of 10 simulations and the interval around them corresponds to the standard deviation. The black dashed line is the theoretical evolution in the absence of $q_{min}$. The bottom left panel zooms into the upper plot around $q_{min}$, showing how the evolution finds equilibrium for small $B$ values. The right panel shows the corresponding evolution of the number of species that are formed in each case, and the black dashed line shows the null expectation for the number of species (Eq.\eqref{eq:nullspecies}). The simulation parameters are $N=1000$, $\mu=0.0025$ and $q_{min}=0.8$.}
    \label{fig:speciesformation}
\end{figure}

Starting with a clonal population, the similarity distribution starts to move towards its equilibrium $q_{eq}$, getting wider at every generation (see Fig.\ref{fig:networks}). When the distribution is close to the threshold $q_{min}$, part of the population is still able to reproduce, while many pairs become incompatible. In a subsequent generation, if the average similarity decreases too much and the distribution is not wide enough, then the number of pairs of individuals that are still able to breed may not be enough to generate another connected community, and new species may appear. Figure \ref{fig:speciesformation} shows the evolution of the average similarity and the formation of species throughout the generations. For small genome sizes the similarity distribution gets ``trapped'' around the threshold $q_{min}$ and evolution does not lead to the formation of species. For larger genome sizes, on the other hand, after a transient region (radiation period) many species are formed and the average similarity keeps following the previous tendency.

We can express this situation as follows: for $\langle q_{\tau}^{\alpha\beta}\rangle\approx q_{min}+\Delta$ sufficiently close to $q_{min}$, speciation may occur if the next generation makes $\langle q_{\tau}^{\alpha\beta}\rangle$ cross the gap, i.e.,
\begin{equation}
    |\langle q_{\tau+1}^{\alpha\beta}\rangle-\langle q_{\tau}^{\alpha\beta}\rangle|>\Delta
\end{equation}
where $\Delta$ is a small quantity that should decrease as $B$ increases. Using Eq.\eqref{eq:mean}  to calculate the step size $|\langle q_{\tau+1}^{\alpha\beta}\rangle-\langle q_{\tau}^{\alpha\beta}\rangle|$, we find
\begin{equation}
    \Delta<\frac{q_{min}/q_{eq}-1}{N-1}\equiv\delta q.\label{eq:delta2}
\end{equation}

The challenge is to calculate $\Delta$ as a function of $B$. As an ansatz, we propose
\begin{equation}
    \Delta=\sqrt{\sigma_B^2}-\lim_{B\rightarrow\infty}\sqrt{\sigma_B^2}\label{eq:ansatz}
\end{equation}
where $\sigma_B^2$ is the variance of the similarity distribution at the generation $\tau$ when $\langle q_{\tau}^{\alpha\beta}\rangle= q_{min}$. In this ansatz, for $B\rightarrow\infty$, $\Delta=0$ and then the transition, according to Eq.\eqref{eq:delta2}, is given by $q_{eq}<q_{min}$, as is it should.

Now, using equations \eqref{eq:sigma}, \eqref{eq:delta2} and \eqref{eq:ansatz}, we find that speciation should happen whenever
\begin{equation}
    B>\frac{\Lambda_2(\tau)}{\delta q^2+2\delta q\sqrt{\Lambda_1(\tau)}}\equiv B_c,
\end{equation}
which defines the critical genome size $B_c$. The values of $\Lambda_1(\tau)$ and $\Lambda_2(\tau)$ are obtained numerically by recursively solving the evolution equations of the system
\begin{equation}
\begin{split}
    \langle q_{t+1}^{\alpha\beta}\rangle&=f_1(\langle q_{t}^{\alpha\beta}\rangle),\\
    \langle q_{t+1}^{\alpha\beta\gamma\delta}\rangle&=f_2(\langle q_{t}^{\alpha\beta}\rangle,\langle q_{t}^{\alpha\beta\gamma\delta}\rangle),\\
    \textrm{Var}(q_{t+1}^{\alpha\beta})&=f_3(\langle q_{t}^{\alpha\beta}\rangle,\langle q_{t}^{\alpha\beta\gamma\delta}\rangle,\textrm{Var}(q_{t}^{\alpha\beta}),\textrm{Cov}(q_{t}^{\alpha\beta\gamma})),\\
    \textrm{Cov}(q_{t+1}^{\alpha\beta})&=f_4(\langle q_{t}^{\alpha\beta}\rangle,\langle q_{t}^{\alpha\beta\gamma\delta}\rangle,\textrm{Var}(q_{t}^{\alpha\beta}),\textrm{Cov}(q_{t}^{\alpha\beta\gamma})),
\end{split}\label{eq:recurrences}
\end{equation}
which are all presented Section \ref{sec:allresults} (equations \eqref{eq:meanAppendix}, \eqref{eq:secondoverlap}, \eqref{eq:variancia} and \eqref{eq:covariancia}). This set of equations gives the exact description of the first two moments of the similarity distribution of the system in the absence of $q_{min}$ and for any set of parameters $(N,\mu,q_{min},B)$.

\section{The heuristic solution and simulations\label{sec:solution}}

\begin{figure}
    \centering
    \includegraphics[width=0.9\linewidth]{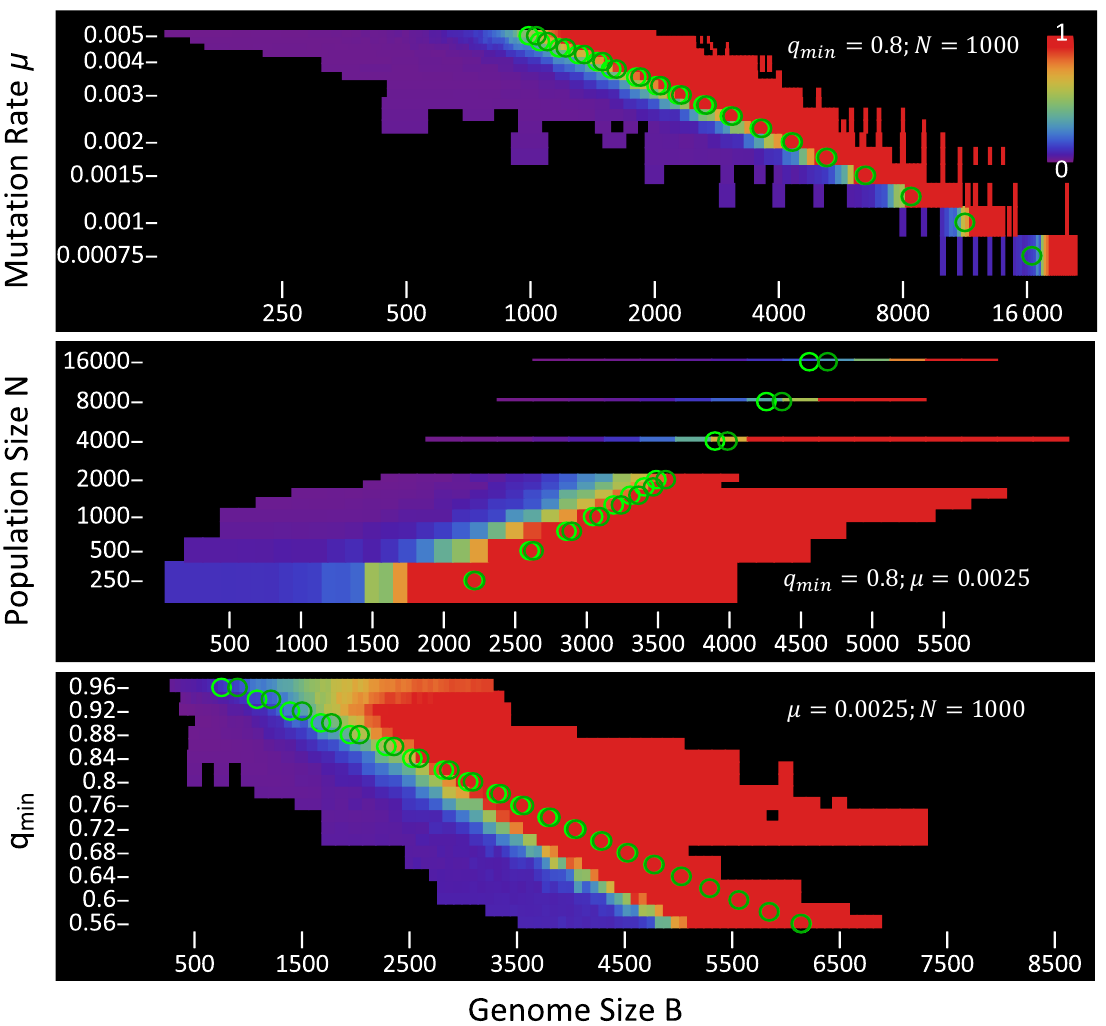}
    \caption{\textbf{Speciation transition in the parameter space and the Heuristic solution.} In the heatmaps, the color code represents the ratio between the observed number of species over a given time range $\overline{S}$ and the expected number of species $S_e$. Whenever this value is greater or equal 1, the point is displayed in red, and it represents a parameter set with speciation, in contrast with the purple region, in which we do not observe speciation. The circles are the numerical solutions for the heuristic critical genome size $B_c$. The light green circles correspond to the solution up to $\lfloor\tau\rfloor$ generations, and the dark green circles to solutions up to $\lceil\tau\rceil$ generations. The simulation parameters are indicated in the figure.}
    \label{fig:heatmaps}
\end{figure}

In order to compare the heuristic solution with computational results of the model, we constructed the heatmaps shown in Figure \ref{fig:heatmaps}. In this figure, for every parameter set $(N,\mu,q_{min},B)$, we ran the Derrida-Higgs dynamics 5 times, starting from a clonal population, for $3\tau$ generations, where $\tau$ is the time it takes for the average similarity to reach $q_{min}$ in the absence of mating restriction, i.e., $\langle q_{\tau}^{\alpha\beta}\rangle=q_{min}$ in equation \eqref{eq:mean},
\begin{equation}
    \tau=\frac{\ln\left[\frac{q_{min}-q_{eq}}{1-q_{eq}}\right]}{\ln\left[\left(1-\frac{1}{N}\right)e^{-4\mu}\right]}.\label{eq:tempo}
\end{equation}
In each simulation, we calculated the average number of species in the last $\tau$ generations. This scheme was chosen in order to try to avoid the radiation period when calculating the number of species but still keeping the simulation times feasible.
The obtained number of species was then averaged over the 5 simulations of the considered parameter set, resulting in $\overline{S}$.

We can estimate the number of species a population would have by calculating the size of a population whose equilibrium $q_{eq}$ is exactly at $q_{min}$,
\begin{equation}
    q_{min}=q_{eq}\Rightarrow N^*=\frac{1}{e^{4\mu}-1}\left(\frac{1}{q_{min}}-1\right).
\end{equation}
Therefore, the number of species a population of size $N$ would show after speciation is
\begin{equation}
    S_e=\frac{N}{N^*}=N(e^{4\mu}-1)\left(\frac{1}{q_{min}}-1\right)^{-1}.\label{eq:nullspecies}
\end{equation}

Computational experiments show that this value underestimates the actual observed number of species, as a result of the asymmetry of the species abundance distribution, leading to species with higher similarity than $q_{min}$ and smaller sizes than $N^*$. On the other hand, $S_e$ is a good scale for the number of species a community can generate, since even when the speciation is slow (when the genome size is not too large) we can still expect it would reach at least $S_e$ species after a sufficiently long time.

Therefore, we consider $\overline{S}/S_e$ a good index for species formation: if $\overline{S}/S_e>1$, then at least the null expected number of species have been formed. The heatmaps of Fig.\ref{fig:heatmaps} plot this ratio in red whenever this index is greater or equal to one. Hence the red region in the parameter set indicates species formation. Whenever this index is close to zero, the color code is closer to purple, indicating no species formation. (The black region correspond to non-simulated parameters).

Because $\tau$ as given by equation \eqref{eq:tempo} is not an integer, to solve equations \eqref{eq:recurrences} up to $\tau$ we may consider its Floor value ($\lfloor\tau\rfloor$) or its Ceiling value ($\lceil\tau\rceil$), which would end in slightly different values for $\Lambda_1(\tau)$ and $\Lambda_2(\tau)$. In Fig.\ref{fig:heatmaps}, for completeness, both values are presented: light green circles for $\lfloor\tau\rfloor$ and dark green circles for $\lceil\tau\rceil$. We notice that, although overestimating the critical value $B_c$, the heuristic solution is in good agreement with the computational results for a wide range of parameters.

\section{Conclusions}
\label{conclusions}

While the formation of species in the absence of gene flow (allopatric speciation) is a well comprehended process, sympatric speciation remains counter-intuitive and lacking a robust set of examples in nature, i.e, examples that would pass the modern tests to assure the sympatric emergence of the species \cite{fitzpatrick2008if,sun2022sympatric,richards2018don}. The Derrida-Higgs model \cite{higgs1991stochastic,higgs1992genetic} shows that sympatric speciation in neutrally evolving communities is, in principle, possible and in this paper we worked on the mathematical details behind the model's theory. The dynamics considers a finite population whose individuals are described by a finite size haploid binary genome evolving under assortative mating and sexual reproduction, without generational overlap. The model is sympatric since it does not present any spatial structure: individuals are not spread over a lattice \cite{de2009global} or separated into demes \cite{manzo1994geographic,princepe2022diversity}. In the sense of Gavrilets \cite{gavrilets2003perspective}, it is a truly sympatric model, representing a situation in the extreme of the ``sympatry--allopatry'' axis. It is noteworthy that a rather similar model was also proposed by Serva and Peliti in the same year (1991), with initial results concerning the fluctuations of the similarity distribution \cite{serva1991statistical}.

In the genetic space, evolution leads to reproductive isolated groups, i.e. different species, depending on the parameters of the model (population size, mutation rate, assortative parameter and genome size). The transition between populations with a single species and many species was already understood for infinite genome sizes, but effects of finite genomes were reported only on 2017 \cite{de2017speciation}, and no theory has been proposed. In this paper, we started to fill this theoretical gap, introducing an analytical description of the evolution of the similarity distributions and a heuristic solution for the speciation transition.

When there are no restrictions to mating, the theory leads to recurrence equations for the moments of the similarity distribution. Notwithstanding, higher order overlap quantities are involved in such calculations, hindering analytical predictions, but the same reasoning applied here can be used to calculate the probability distributions involved in such overlap quantities.  As far as the authors know, all results concerning finite number of alleles are new. We have shown, in particular, that the evolution of the average similarity, that had been obtained in the limit of infinite genome size, is actualy independent of genome size and works for all values of $B$. We also calculated the recurrence equations for the variance of the distribution, whose functional dependence on the genome size is crucial to understand the speciation transition. 

When mating restrictions are imposed, we used a heuristic approach to describe the speciation transition, i.e., to explain why only under certain conditions the population breaks into non connected groups of individuals. When speciation happens, the average similarity of the population gets smaller than a connected group can sustain. We propose that such threshold is measured as a nontrivial function of the variance of the similarity distribution, an ansatz we used to numerically calculate the critical genome size that enables species formation in the system. Although approximate, our solution is the first to describe this process and it is in good agreement with simulation results. The fact that our procedure overestimates the critical genome size indicates that our heuristic solution may be too restrictive.

Despite its simplicity, the Derrida-Higgs model is an interesting framework for studying evolutionary processes, and several extensions of this model have been proposed in \cite{de2009global,costa2019signatures,princepe2021modeling,princepe2022mito}. In spatial structured versions of the model, the critical genome size for speciation is much smaller than in the sympatric case, as a consequence of the rather reduced number of individuals in the reproduction range \cite{de2017speciation}. In Princepe and de Aguiar \cite{princepe2021modeling}, the inclusion of a second binary chain in spatial models was used to study the coevolution between the nuclear and the mitochondrial DNA, showing how wide is the range of applications of the Derrida-Higgs model, and understanding its basic foundations was our goal in this paper. Future work in the field may include ecological interactions in the model and investigate how the interplay of ecological and evolutionary time-scales may affect the thresholds of diversification.

\begin{acknowledgments}

V.M.M. was partially supported by the grant 140728/2019-8, of ``Conselho Nacional de Desenvolvimento Cient\'ifico e Tecnol\'ogico'', CNPq,  and by the grants 2019/13341-7 and 2021/12509-1, São Paulo Research Foundation (FAPESP). V.M.M. also acknowledges the Abdus Salam International Centre for Theoretical Physics (ICTP) for partially hosting the researcher during the development of this project.

V.M.M. is currently affiliated to Aix Marseille Univ, CNRS, Laboratoire de Chimie Bactérienne
(UMR7283), Turing Centre for Living Systems, IMM, IM2B, Marseille, France.

\end{acknowledgments}

\bibliography{references.bib}


\pagebreak
\widetext
\begin{center}
	\textbf{\large Supplementary Material of:\\``The transition to speciation in the finite genome Derrida-Higgs model: a heuristic solution''}\\
	Vitor M. Marquioni$^1$, Marcus A. M. de Aguiar$^1$\\
	\textit{$^1$Institute of Physics Gleb Wataghin, University of Campinas, Campinas, SP, Brazil}
\end{center}

\setcounter{equation}{0}
\setcounter{section}{0}
\setcounter{figure}{0}
\setcounter{table}{0}
\setcounter{page}{1}
\makeatletter
\renewcommand{\theequation}{S\arabic{equation}}
\renewcommand{\thefigure}{S\arabic{figure}}
\renewcommand{\bibnumfmt}[1]{[S#1]}
\renewcommand{\citenumfont}[1]{S#1}

\section{\label{sec:introduction}Introduction}


In this supplementary material, we carry on the calculations behind the results summed up in the section V of the main text. We start at the end of the calculation of the similarity distribution, which is developed in the main text and calculate its first and second moments. We calculate the covariance of similarities between different pairs of individuals and all other important quantities that appear. Higher-order overlaps are introduced and some of their properties are discussed at the end.

\section{The similarity distribution}

\subsection{Probability of drawing a pair of parents}

In Section III of the main text 
we wrote down the probability of drawing a given pair of individuals $(p_1,p_2)$ from generation $t$ to be the parents of the individual $\alpha$ in generation $t+1$. Here we calculate this probability explicitly. We first separate it in two parts, first supposing that $p_1$ is the focal individual, and then supposing that $p_2$ is yhr focal individual,
\begin{align}
	\mathcal{P}((p_1,p_2))=&\mathcal{P}((p_1,p_2)|p_1\textrm{ is focal})\mathcal{P}(p_1\textrm{ is focal})+\mathcal{P}((p_1,p_2)|p_2\textrm{ is focal})\mathcal{P}(p_2\textrm{ is focal}).\nonumber
\end{align}
The probability that any individual is drawn as focal is $1/N$, since there are no fitness differences between the any individual. Let us consider the adjacency matrix $\mathbb{A}$ whose elements $A_{ij}$ are equal to 1 when $q^{ij}\ge q_{min}$ and 0 otherwise (as also for $i=j$). The community then describes a network whose nodes are connected if they are similar enough to mate. If the degree of the node $i$, given by $N_i=\sum_jA_{ij}$, counts how many individuals are connected to $i$, then given that $i$ is a focal individual, any other individual $j$ can be its mating partner with probability $A_{ij}/N_{i}$. Therefore,
\begin{align}
	\mathcal{P}((p_1,p_2))=\frac{A_{p_1p_2}}{N}\left(\frac{1}{N_{p_1}}+\frac{1}{N_{p_2}}\right),
\end{align}
which is the Eq.(13) in the main text.

Also, from the main text, we can write the similarity probability distribution as
\begin{align}
	\mathcal{P}(q_{t+1}^{\alpha\beta})=&\frac{1}{N^2}
	\sum_{\mathbf{S}^{\alpha},\mathbf{S}^{\beta}}
	\sum_{(p_1,p_2)}\sum_{(p'_1,p'_2)}
	\mathcal{P}(q_{t+1}^{\alpha\beta}|\mathbf{S}^{\alpha},\mathbf{S}^{\beta})
	\mathcal{P}(\mathbf{S}^{\alpha}|(p_1,p_2)) \mathcal{P}(\mathbf{S}^{\beta}|(p_1',p_2'))\mathcal{P}((p_1,p_2))\mathcal{P}((p_1',p_2')),
\end{align}
where each term has already been calculated. As a reminder of the notation: $(p_1,p_2)$ are the parents of $\alpha$ and $(p_1',p_2')$ are the parents of $\beta$. Therefore, everything that is related to $p_1,p_2,p_1'$ or $p_2'$ is calculated at time $t$, and everything that refers to $\alpha$ and $\beta$ is calculated at time $t+1$.

Hence,
\begin{align}
	\mathcal{P}(q_{t+1}^{\alpha\beta})=&\frac{1}{N^2}\sum_{\mathbf{S}^{\alpha},\mathbf{S}^{\beta}}\sum_{(p_1,p_2)}\sum_{(p'_1,p'_2)}\delta\left(q_{t+1}^{\alpha\beta},\frac{\mathbf{S}^{\alpha}\cdot\mathbf{S}^{\beta}}{B}\right)A_{p_1p_2}A_{p'_1p'_2}\left(\frac{1}{N_{p_1}}+\frac{1}{N_{p_2}}\right)\left(\frac{1}{N_{p_1'}}+\frac{1}{N_{p_2'}}\right) \times\nonumber\\\times&\prod_{i=1}^B\left[\frac{1}{2}+\frac{s_{i,t+1}^{\alpha} e^{-2\mu}}{4}\left(s_{i,t}^{p_1}+s_{i,t}^{p_2}\right)\right]\left[\frac{1}{2}+\frac{s_{i,t+1}^{\beta} e^{-2\mu}}{4}\left(s_{i,t}^{p_1'}+s_{i,t}^{p_2'}\right)\right],\label{eq:probMiddleStep}
\end{align}
which is still not the same equation we see in the main text.
Notice that here the sums regarding the parents are performed over the pairs of individuals, not over the individuals themselves. We can change it by noticing that the sums are the same if one interchanges $p_1$ and $p_2$,
\[
\sum_{(p_1,p_2)}\longrightarrow\frac{1}{2}\sum_{p_1,p_2}.
\]
The case $p_1=p_2$ could be a problem, but it is automatically solved, once $A_{p_1p_1}=0$. Obviously, the same works for the sums over $(p_1',p_2')$. Moreover,
\begin{align}
	&\sum_{p_1,p_2}\frac{A_{p_1p_2}}{N_{p_2}}\prod_{i=1}^B\left[\frac{1}{2}+\frac{s_{i,t+1}^{\alpha} e^{-2\mu}}{4}\left(s_{i,t}^{p_1}+s_{i,t}^{p_2}\right)\right]\nonumber\\=&\sum_{p_2,p_1}\frac{A_{p_2p_1}}{N_{p_2}}\prod_{i=1}^B\left[\frac{1}{2}+\frac{s_{i,t+1}^{\alpha} e^{-2\mu}}{4}\left(s_{i,t}^{p_2}+s_{i,t}^{p_1}\right)\right]\nonumber\\=&\sum_{p_1,p_2}\frac{A_{p_1p_2}}{N_{p_1}}\prod_{i=1}^B\left[\frac{1}{2}+\frac{s_{i,t+1}^{\alpha} e^{-2\mu}}{4}\left(s_{i,t}^{p_1}+s_{i,t}^{p_2}\right)\right].\nonumber
\end{align}
In the second line we have inverted the order of the sums and used that $A_{p_1p_2}=A_{p_2p_1}$, while in the third line, we have changed the index names, $p_1\rightarrow p_2$ and $p_2\rightarrow p_1$. Thus,
\begin{align}
	&\sum_{p_1,p_2}A_{p_1p_2}\left(\frac{1}{N_{p_1}}+\frac{1}{N_{p_2}}\right)\prod_{i=1}^B\left[\frac{1}{2}+\frac{s_{i,t+1}^{\alpha} e^{-2\mu}}{4}\left(s_{i,t}^{p_1}+s_{i,t}^{p_2}\right)\right]\nonumber\\=&2\sum_{p_1,p_2}\frac{A_{p_1p_2}}{N_{p_1}}\prod_{i=1}^B\left[\frac{1}{2}+\frac{s_{i,t+1}^{\alpha} e^{-2\mu}}{4}\left(s_{i,t}^{p_1}+s_{i,t}^{p_2}\right)\right]\nonumber
\end{align}
in such a way that $\mathcal{P}(q_{t+1}^{\alpha\beta})$ can be written as
\begin{align}
	\mathcal{P}(q_{t+1}^{\alpha\beta})=&\frac{1}{N^2}\sum_{\mathbf{S}^{\alpha},\mathbf{S}^{\beta}}\sum_{p_1,p_2}\sum_{p'_1,p'_2}\delta\left(q_{t+1}^{\alpha\beta},\frac{\mathbf{S}^{\alpha}\cdot\mathbf{S}^{\beta}}{B}\right)\frac{A_{p_1p_2}A_{p'_1p'_2}}{N_{p_1}N_{p_1'}} \times\nonumber\\\times&\prod_{i=1}^B\left[\frac{1}{2}+\frac{s_{i,t+1}^{\alpha} e^{-2\mu}}{4}\left(s_{i,t}^{p_1}+s_{i,t}^{p_2}\right)\right]\left[\frac{1}{2}+\frac{s_{i,t+1}^{\beta} e^{-2\mu}}{4}\left(s_{i,t}^{p_1'}+s_{i,t}^{p_2'}\right)\right],
	\label{eq:distribuicao}
\end{align}
which is Eq. 14 in the main text. This equation allows us now to calculate the similarity probability distribution at time $t+1$ given all the genetic information about the system at time $t$.

\subsection{The evolution of the Mean Similarity}

Once we have calculated the similarity distribution, we can calculate its moments.
Let us start by calculating the expected value of $q_{t+1}^{\alpha\beta}$, given by
\[
\mathbb{E}(q_{t+1}^{\alpha\beta})=\sum_{q}q\mathcal{P}(q_{t+1}^{\alpha\beta}=q).
\]
Because of the $\delta$ function, the sum over $q$ is easy to perform and the term $q\delta(q,\mathbf{S}^{\alpha}\cdot\mathbf{S}^{\beta}/B)$ changes to $\mathbf{S}^{\alpha}\cdot\mathbf{S}^{\beta}/B$, decoupling the sums over $\mathbf{S}^{\alpha}$ and $\mathbf{S}^{\beta}$ from the others,
\begin{align}
	\mathbb{E}(q_{t+1}^{\alpha\beta})=\frac{1}{N^2}\frac{1}{B}\sum_{\mathbf{S}^{\alpha},\mathbf{S}^{\beta}}\sum_{p_1,p_2}\sum_{p'_1,p'_2}\sum_{j=1}^Bs_j^{\alpha}s_j^{\beta}\frac{A_{p_1p_2}A_{p'_1p'_2}}{N_{p_1}N_{p'_1}}\prod_{i=1}^BF_{i,t}(\alpha,p_1,p_2)\prod_{i=1}^BF_{i,t}(\beta,p'_1,p'_2)\nonumber
\end{align}
where we have defined
\begin{equation}
	F_{i,t}(\gamma,a,b)\equiv\frac{1}{2}+\frac{s_{i,t+1}^{\gamma}e^{-2\mu}}{4}\left(s_{i,t}^{a}+s_{i,t}^{b}\right)=\mathcal{P}(s_{i,t+1}^{\gamma}|(a,b))
\end{equation}
the probability of the allele $s_{i,t+1}^{\gamma}$ of $\gamma$ given the alleles of its parents $a$ and $b$.
Rearranging the sums,
\begin{align}
	\mathbb{E}(q_{t+1}^{\alpha\beta})=\frac{1}{N^2}\frac{1}{B}\sum_{p_1,p_2}\sum_{p'_1,p'_2}\frac{A_{p_1p_2}A_{p'_1p'_2}}{N_{p_1}N_{p'_1}}\sum_j\left(\sum_{\mathbf{S}^{\alpha}}s_j^{\alpha}\prod_{i=1}^BF_{i,t}(\alpha,p_1,p_2)\right)\left(\sum_{\mathbf{S}^{\beta}}s_j^{\beta}\prod_{i=1}^BF_{i,t}(\beta,p'_1,p'_2)\right)\nonumber
\end{align}

The terms in parenthesis can be calculated as follows,
\begin{align}
	\sum_{\mathbf{S}^{\alpha}}s_j^{\alpha}\prod_{i=1}^BF_{i,t}(\alpha,p_1,p_2)&=\sum_{s_1^{\alpha},\dots,s_B^{\alpha}}s_j^{\alpha}\prod_{i=1}^BF_{i,t}(\alpha,p_1,p_2)\nonumber\\&=\sum_{s_1^{\alpha},\dots,s_B^{\alpha}}F_{1,t}\dots F_{j-1,t}s_{j,t}^{\alpha}F_{j,t}F_{j+1,t}\dots F_{B,t}\nonumber\\&=\left(\sum_{s_1^{\alpha}}F_{1,t}\right)\dots \left(\sum_{s_j^{\alpha}}s_j^{\alpha}F_{j,t}\right)\dots\left(\sum_{s_{B,t}}F_{B,t}\right)\nonumber
\end{align}
These last terms are easily calculated,
\begin{align}
	\sum_{s_k^{\alpha}}F_{k,t}=\sum_{s_k^{\alpha}=\pm1}F_{k,t}=1,\nonumber
\end{align}
and
\begin{align}
	\sum_{s_k^{\alpha}}s_k^{\alpha}F_{k,t}=\sum_{s_k^{\alpha}=\pm1}s_k^{\alpha}F_{k,t}=\frac{e^{-2\mu}}{2}(s_k^{p_1}+s_k^{p_2}).\nonumber
\end{align}
Thus,
\begin{align}
	\sum_{\mathbf{S}^{\alpha}}s_j^{\alpha}\prod_{i=1}^BF_{i,t}(\alpha,p_1,p_2)&=\frac{e^{-2\mu}}{2}(s_j^{p_1}+s_j^{p_2}).
\end{align}

Now, the expected value is given by
\begin{align}
	\mathbb{E}(q_{t+1}^{\alpha\beta})=\frac{1}{N^2}\frac{1}{B}\sum_{p_1,p_2}\sum_{p'_1,p'_2}\frac{A_{p_1p_2}A_{p'_1p'_2}}{N_{p_1}N_{p'_1}}\sum_{j=1}^B\frac{e^{-4\mu}}{4}(s_j^{p_1}+s_j^{p_2})(s_j^{p_1'}+s_j^{p_2'}),\nonumber
\end{align}
and remembering that the terms relative to the parents are calculated at time $t$, we end up with
\begin{align}
	\mathbb{E}(q_{t+1}^{\alpha\beta})=\frac{e^{-4\mu}}{4N^2}\sum_{p_1,p_2}\sum_{p'_1,p'_2}\frac{A_{p_1p_2}A_{p'_1p'_2}}{N_{p_1}N_{p'_1}}\left(q_t^{p_1p_1'}+q_t^{p_1p_2'}+q_t^{p_2p_1'}+q_t^{p_2p_2'}\right).
	\label{eq:mediageral}
\end{align}

\subsection{The mean without assortative reproduction}

In the general case, equation \eqref{eq:mediageral} is not easy to deal with, but in the absence of assortative reproduction, i.e., without $q_{min}$ (or $q_{min}=-1$), this equation is quite simple. Since there are no limitations to the reproduction, any individual can mate with any other and then this case is equivalent to a complete network without loops: $A_{p_1p_2}=A_{p_1'p_2'}=1$ for any $p_1\ne p_2$ and $p_1'\ne p_2'$, and $N_{p_1}=N_{p_1'}=N-1$. Thus,
\begin{align}
	\mathbb{E}(q_{t+1}^{\alpha\beta})=\frac{e^{-4\mu}}{4N^2(N-1)^2}\sum_{p_1,p_2}\sum_{p'_1,p'_2}A_{p_1p_2}A_{p'_1p'_2}\left(q_t^{p_1p_1'}+q_t^{p_1p_2'}+q_t^{p_2p_1'}+q_t^{p_2p_2'}\right).
	\nonumber
\end{align}
And now, by changing the order and indexes names,
\begin{align}
	\mathbb{E}(q_{t+1}^{\alpha\beta})&=\frac{e^{-4\mu}}{N^2(N-1)^2}\sum_{p_1,p_2}\sum_{p'_1,p'_2}A_{p_1p_2}A_{p'_1p'_2}q_t^{p_1p_1'}\nonumber\\&=\frac{e^{-4\mu}}{N^2(N-1)^2}\sum_{p_1,p_1'}\left(\sum_{p_2}A_{p_1p_2}\right)\left(\sum_{p_2'}A_{p'_1p'_2}\right)q_t^{p_1p_1'}\nonumber\\&=\frac{e^{-4\mu}}{N^2}\sum_{p_1,p_1'}q_t^{p_1p_1'},
	\label{eq:reqmediasemqmin}
\end{align}
since $\sum_jA_{ij}=N_j=N-1$. The expression above calculates the expected value of the similarity distribution at time $t+1$ given the similarity values at time $t$ and can be simplified as
\begin{align}
	\mathbb{E}(q_{t+1}^{\alpha\beta})&=\frac{e^{-4\mu}}{N^2}\sum_{p_1,p_1'} q_t^{p_1p_1'}\nonumber\\&=\frac{e^{-4\mu}}{N^2}\left(N+\sum_{p_1\ne p_1'}q_t^{p_1p_1'}\right).
	\nonumber
\end{align}

In this equation, we can recognize the average similarity within the population, given by $\langle q_t^{\alpha\beta}\rangle_P=\frac{1}{N(N-1)}\sum_{p_1\ne p_1'}q_t^{p_1p_1'}$, then
\begin{align}
	\mathbb{E}(q_{t+1}^{\alpha\beta})= \frac{e^{-4\mu}}{N^2}\left(N+N(N-1)\langle q_t^{\alpha\beta}\rangle_P\right)=e^{-4\mu}\left[\frac{1}{N}+\left(1-\frac{1}{N}\right)\langle q_t^{\alpha\beta}\rangle_P\right].\label{eq:eqmedia1}
\end{align}
This equation is interesting because it relates the expected value for the similarity between a pair of individuals in the next step with the observed average similarity of the present population. On the other hand, this is not a recurrence equation. In order to find it, one can take the ensemble average of both sides by averaging over all the possible trajectories $\mathcal{T}_t$ up to time $t$, and once the equation is linear, it is possible to write
\begin{align}
	\langle q_{t+1}^{\alpha\beta}\rangle=e^{-4\mu}\left[\frac{1}{N}+\left(1-\frac{1}{N}\right)\langle q_t^{\alpha\beta}\rangle\right],\label{eq:eqmedia2}
\end{align}
which is equation (16) from the main text.

\subsection{The evolution of the variance}

\begingroup
\allowdisplaybreaks

Now, in order to calculate the variance of the distribution,
\begin{equation}
	\textrm{Var}(q_{t+1}^{\alpha\beta})=\mathbb{E}((q_{t+1}^{\alpha\beta})^2)-\mathbb{E}(q_{t+1}^{\alpha\beta})^2,
\end{equation}
we must calculate the second moment
\begin{equation}
	\mathbb{E}((q_{t+1}^{\alpha\beta})^2)= \sum_{q}q^2\mathcal{P}(q_{t+1}^{\alpha\beta}=q).
\end{equation}
It is given by
\begin{align}
	\mathbb{E}((q_{t+1}^{\alpha\beta})^2)&=\frac{1}{N^2}\sum_{\mathbf{S}^{\alpha},\mathbf{S}^{\beta}}\sum_{p_1,p_2}\sum_{p'_1,p'_2}\left(\frac{\mathbf{S}^{\alpha}\cdot\mathbf{S}^{\beta}}{B}\right)^2\frac{A_{p_1p_2}A_{p'_1p'_2}}{N_{p_1}N_{p'_1}}\prod_{i=1}^BF_{i,t}(\alpha,p_1,p_2)F_{i,t}(\beta,p'_1,p'_2)\nonumber\\
	&=\frac{1}{N^2B^2}\sum_{\mathbf{S}^{\alpha},\mathbf{S}^{\beta}}\sum_{p_1,p_2}\sum_{p'_1,p'_2}\left(\sum_{j,k}s_j^{\alpha}s_j^{\beta}s_k^{\alpha}s_k^{\beta}\right)\frac{A_{p_1p_2}A_{p'_1p'_2}}{N_{p_1}N_{p'_1}}\prod_{i=1}^BF_{i,t}(\alpha,p_1,p_2)F_{i,t}(\beta,p'_1,p'_2)\nonumber\\
	&=\frac{1}{N^2B^2}\sum_{p_1,p_2}\sum_{p'_1,p'_2}\frac{A_{p_1p_2}A_{p'_1p'_2}}{N_{p_1}N_{p'_1}}\nonumber\\
	&\times\sum_{\mathbf{S}^{\alpha},\mathbf{S}^{\beta}}\left(\sum_{j=1}^Bs_j^{\alpha}s_j^{\beta}s_j^{\alpha}s_j^{\beta}+\sum_{j\ne k}s_j^{\alpha}s_j^{\beta}s_k^{\alpha}s_k^{\beta}\right)\prod_{i=1}^BF_{i,t}(\alpha,p_1,p_2)F_{i,t}(\beta,p'_1,p'_2)\nonumber\\
	&=\frac{1}{N^2B^2}\sum_{p_1,p_2}\sum_{p'_1,p'_2}\frac{A_{p_1p_2}A_{p'_1p'_2}}{N_{p_1}N_{p'_1}}\nonumber\\
	&\times\sum_{\mathbf{S}^{\alpha},\mathbf{S}^{\beta}}\left[B\prod_{i=1}^BF_{i,t}(\alpha)F_{i,t}(\beta)+\sum_{j\ne k}\left(s_j^{\alpha}s_k^{\alpha}\prod_{i=1}^BF_{i,t}(\alpha)\right)\left(s_j^{\beta}s_k^{\beta}\prod_{i=1}^BF_{i,t}(\beta)\right)\right].\nonumber\\
\end{align}

In the expression above, the first term in brackets, after summing on $\mathbf{S}^{\alpha}$ and $\mathbf{S}^{\beta}$, equals to $B$ (as we have already calculated). The second term is also easy to calculate if we follow the same procedure for the expected value $\mathbb{E}(q_{t+1}^{\alpha\beta})$, the difference now is that two terms in the product do not equal to 1, instead of only one,
\begin{align}
	&\sum_{\mathbf{S}^{\alpha},\mathbf{S}^{\beta}}\sum_{j\ne k}\left(s_j^{\alpha}s_k^{\alpha}\prod_{i=1}^BF_{i,t}(\alpha)\right)\left(s_j^{\beta}s_k^{\beta}\prod_{i=1}^BF_{i,t}(\beta)\right)\nonumber\\
	&=\sum_{j\ne k}\frac{e^{-8\mu}}{16}(s_j^{p_1}+s_j^{p_2})(s_j^{p_1'}+s_j^{p_2'})(s_k^{p_1}+s_k^{p_2})(s_k^{p_1'}+s_k^{p_2'})\nonumber\\
	&=\frac{e^{-8\mu}}{16}\sum_{j=1}^B(s_j^{p_1}+s_j^{p_2})(s_j^{p_1'}+s_j^{p_2'})\sum_{k=1,k\ne j}^B(s_k^{p_1}+s_k^{p_2})(s_k^{p_1'}+s_k^{p_2'})\nonumber\\
	&=\frac{e^{-8\mu}}{16}\sum_{j=1}^B(s_j^{p_1}+s_j^{p_2})(s_j^{p_1'}+s_j^{p_2'})\left[-(s_j^{p_1}+s_j^{p_2})(s_j^{p_1'}+s_j^{p_2'})+\sum_{k=1}^B(s_k^{p_1}+s_k^{p_2})(s_k^{p_1'}+s_k^{p_2'})\right]\nonumber\\
	&=\frac{e^{-8\mu}}{16}\left[B^2(q_t^{p_1p_1'}+q_t^{p_1p_2'}+q_t^{p_2p_1'}+q_t^{p_2p_2'})^2-\sum_{j=1}^B(s_j^{p_1}+s_j^{p_2})^2(s_j^{p_1'}+s_j^{p_2'})^2\right]\nonumber\\
	&=\frac{e^{-8\mu}}{16}\left[B(q_t^{p_1p_1'}+q_t^{p_1p_2'}+q_t^{p_2p_1'}+q_t^{p_2p_2'})\right]^2-\frac{e^{-8\mu}}{4}\left[B+Bq_t^{p_1p_2}+Bq_t^{p_1'p_2'}+\sum_{j=1}^Bs_j^{p_1}s_j^{p_2}s_j^{p_1'}s_j^{p_2'}\right].
\end{align}

Defining the \emph{second order overlap} among individuals $\alpha$, $\beta$, $\gamma$ and $\delta$ as
\begin{equation}
	q^{\alpha\beta\gamma\delta}=\frac{1}{B}\sum_{i=1}^Bs_i^{\alpha}s_i^{\beta}s_i^{\gamma}s_i^{\delta}
\end{equation}
we can write,
\begin{align}
	\mathbb{E}((q_{t+1}^{\alpha\beta})^2) &=\frac{1}{N^2}\sum_{p_1,p_2}\sum_{p'_1,p'_2}\frac{A_{p_1p_2}A_{p'_1p'_2}}{N_{p_1}N_{p'_1}}\left[\frac{1}{B}+\frac{e^{-8\mu}}{16}\left[(q_t^{p_1p_1'}+q_t^{p_1p_2'}+q_t^{p_2p_1'}+q_t^{p_2p_2'})\right]^2\right.\nonumber\\
	&\left.-\frac{e^{-8\mu}}{4B}\left(1+q_t^{p_1p_2}+q_t^{p_1'p_2'}+q_t^{p_1p_2p_1'p_2'}\right)\right],\label{eq:secondmoment1}
\end{align}
which is equation (17) from the main text.

\subsection{The variance without assortative reproduction}

As for the mean similarity, the equation for the variance is not easy to treat in the general case, but when there is no $q_{min}$, due to the network structure, it is possible to calculate it. However, it is not as simple as for the mean and we must treat the sums very carefully,
\begin{align}
	\mathbb{E}((q_{t+1}^{\alpha\beta})^2) &=\frac{1}{B}-\frac{e^{-8\mu}}{4B}\left(1+\frac{2}{N}\sum_{p_1,p_2}\frac{A_{p_1p_2}}{(N-1)}q_t^{p_1p_2}+\frac{1}{N^2}\sum_{p_1,p_2}\sum_{p'_1,p'_2}\frac{A_{p_1p_2}A_{p'_1p'_2}}{(N-1)^2}q_t^{p_1p_2p_1'p_2'}\right)\nonumber\\
	&+\frac{e^{-8\mu}}{16N^2}\sum_{p_1,p_2}\sum_{p'_1,p'_2}\frac{A_{p_1p_2}A_{p'_1p'_2}}{(N-1)^2}\left((q_t^{p_1p_1'})^2+(q_t^{p_1p_2'})^2+(q_t^{p_2p_1'})^2+(q_t^{p_2p_2'})^2\right.\nonumber\\
	&\left.+2q_t^{p_1p_1'}q_t^{p_1p_2'}+2q_t^{p_1p_1'}q_t^{p_2p_1'}+2q_t^{p_1p_1'}q_t^{p_2p_2'}+2q_t^{p_1p_2'}q_t^{p_2p_1'}+2q_t^{p_1p_2'}q_t^{p_2p_2'}+2q_t^{p_2p_1'}q_t^{p_2p_2'}\right)\nonumber\\
	&=\frac{1}{B}-\frac{e^{-8\mu}}{4B}\left(1+\frac{2}{N}\sum_{p_1,p_2}\frac{A_{p_1p_2}}{(N-1)}q_t^{p_1p_2}+\frac{1}{N^2}\sum_{p_1,p_2}\sum_{p'_1,p'_2}\frac{A_{p_1p_2}A_{p'_1p'_2}}{(N-1)^2}q_t^{p_1p_2p_1'p_2'}\right)\nonumber\\
	&+\frac{e^{-8\mu}}{4N^2}\left(\sum_{p_1,p'_1}(q_t^{p_1p_1'})^2+2\sum_{p_1}\sum_{p'_1,p'_2}\frac{A_{p'_1p'_2}}{(N-1)}q_t^{p_1p_1'}q_t^{p_1p_2'}+\sum_{p_1,p_2}\sum_{p'_1,p'_2}\frac{A_{p_1p_2}A_{p'_1p'_2}}{(N-1)^2}q_t^{p_1p_1'}q_t^{p_2p_2'}\right).
\end{align}
The double sums are easy to calculate,
\begin{align}
	\mathbb{E}((q_{t+1}^{\alpha\beta})^2)
	&=\frac{1}{B}-\frac{e^{-8\mu}}{4B}\left(1+2\langle q_t^{p_1p_2}\rangle_P+\frac{1}{N^2}\sum_{p_1,p_2}\sum_{p'_1,p'_2}\frac{A_{p_1p_2}A_{p'_1p'_2}}{(N-1)^2}q_t^{p_1p_2p_1'p_2'}\right)\nonumber\\
	&+\frac{e^{-8\mu}}{4N^2}\left(N+N(N-1)\langle(q_t^{p_1p_1'})^2\rangle_P+2\sum_{p_1}\sum_{p'_1,p'_2}\frac{A_{p'_1p'_2}}{(N-1)}q_t^{p_1p_1'}q_t^{p_1p_2'}\right.\nonumber\\
	&\left.+\sum_{p_1,p_2}\sum_{p'_1,p'_2}\frac{A_{p_1p_2}A_{p'_1p'_2}}{(N-1)^2}q_t^{p_1p_1'}q_t^{p_2p_2'}\right),
\end{align}
where we have used that $\sum_{p_1,p_2}A_{p_1p_2}q_t^{p_1p_2}=N(N-1)\langle q_t^{p_1p_2}\rangle_P$ and $\sum_{p_1,p_1'}(q_t^{p_1p_2})^2=N+N(N-1)\langle (q_t^{p_1p_2})^2\rangle_P$. However, the sums over 3 and 4 indexes are not so simple. The best way to deal with them is to open the sums in all possible combinations of indexes as
\begin{align}
	\sum_{p_1,p_1',p_2'}&f(p_1,p_1',p_2')=\nonumber\\
	&\sum_{p_1}\left[\sum_{p_1'=p_1}\sum_{p_2'=p_1}+\sum_{p_1'=p_1}\sum_{p_2'\ne p_1}+\sum_{p_1'\ne p_1}\sum_{p_2'=p_1}+\sum_{p_1'\ne p_1}\sum_{p_2'=p_1'}+\sum_{p_1'\ne p_1}\sum_{p_2'\ne p_1,p_1'}\right]f(p_1,p_1',p_2')\label{eq:soma3}
\end{align}
and
\begin{align}
	\sum_{p_1,p_2,p_1',p_2'}&f(p_1,p_2,p_1',p_2')=\nonumber\\
	&\sum_{p_1}\left[\sum_{p_2=p_1}\sum_{p_1'=p_1}\sum_{p_2'=p_1}+\sum_{p_2=p_1}\sum_{p_1'=p_1}\sum_{p_2'\ne p_1}+\sum_{p_2=p_1}\sum_{p_1'\ne p_1}\sum_{p_2'=p_1}+\sum_{p_2=p_1}\sum_{p_1'\ne p_1}\sum_{p_2'=p_1'}\right.\nonumber\\
	&+\left.\sum_{p_2=p_1}\sum_{p_1'\ne p_1}\sum_{p_2'\ne p_1,p_1'}+\sum_{p_2\ne p_1}\sum_{p_1'=p_1}\sum_{p_2'=p_1}+\sum_{p_2\ne p_1}\sum_{p_1'=p_1}\sum_{p_2'=p_2}+\sum_{p_2\ne p_1}\sum_{p_1'=p_1}\sum_{p_2'\ne p_1,p_2}\right.\nonumber\\
	&+\left.\sum_{p_2\ne p_1}\sum_{p_1'= p_2}\sum_{p_2'=p_1}+\sum_{p_2\ne p_1}\sum_{p_1'=p_2}\sum_{p_2'=p_2}+\sum_{p_2\ne p_1}\sum_{p_1'=p_2}\sum_{p_2'\ne p_1,p_2}+\sum_{p_2\ne p_1}\sum_{p_1'\ne p_1,p_2}\sum_{p_2'= p_1}\right.\nonumber\\
	&+\left.\sum_{p_2\ne p_1}\sum_{p_1'\ne p_1,p_2}\sum_{p_2'=p_2}+\sum_{p_2\ne p_1}\sum_{p_1'\ne p_1,p_2}\sum_{p_2'=p_1'}+\sum_{p_2\ne p_1}\sum_{p_1'\ne p_1,p_2}\sum_{p_2'\ne p_1,p_2,p_1'}\right]f(p_1,p_2,p_1',p_2').\label{eq:soma4}
\end{align}

With these manipulations, the complete expression for each sum term is given by
\begin{align}
	\sum_{p_1,p_2}\sum_{p'_1,p'_2}A_{p_1p_2}A_{p'_1p'_2}q_t^{p_1p_2p_1'p_2'}&=2N(N-1)+4N(N-1)(N-2)\langle q_t^{p_1p'_1}\rangle_P\nonumber\\
	&+N(N-1)(N-2)(N-3)\langle q_t^{p_1p_2p'_1p'_2}\rangle_P\label{eq:eqaux1}
\end{align}
\begin{equation}
	\sum_{p_1}\sum_{p'_1,p'_2}A_{p'_1p'_2}q_t^{p_1p_1'}q_t^{p_1p_2'}=2N(N-1)\langle q_t^{p_1p'_1}\rangle_P+N(N-1)(N-2)\langle q_t^{p_1p'_1}q_t^{p_1p'_2}\rangle_P
\end{equation}
\begin{align}
	\sum_{p_1,p_2}\sum_{p'_1,p'_2}A_{p_1p_2}A_{p'_1p'_2}q_t^{p_1p_1'}q_t^{p_2p_2'}&=N(N-1)+N(N-1)\langle (q_t^{p_1p'_1})^2\rangle_P\nonumber\\
	&+2N(N-1)(N-2)\left(\langle q_t^{p_1p'_1}\rangle_P+\langle q_t^{p_1p'_1}q_t^{p_1p'_2}\rangle_P\right)\nonumber\\
	&+N(N-1)(N-2)(N-3)\langle q_t^{p_1p'_1}q_t^{p_2p'_2}\rangle_P,
\end{align}
where the expressions for population averages are
\begin{equation}
	\langle q_t^{ij}\rangle_P=\frac{1}{N(N-1)}\sum_{p_1}\sum_{p_1'\ne p_1}q_t^{p_1p_1'},
\end{equation}
\begin{equation}
	\langle(q_t^{ij})^2\rangle_P=\frac{1}{N(N-1)}\sum_{p_1}\sum_{p_1'\ne p_1}(q_t^{p_1p_1'})^2,
\end{equation}
\begin{equation}
	\langle q_t^{ij}q_t^{kj}\rangle_P=\frac{1}{N(N-1)(N-2)}\sum_{p_1}\sum_{p_1'\ne p_1}\sum_{p_2'\ne p_1,p_1'}q_t^{p_1p_1'}q_t^{p_1p_2'},
\end{equation}
\begin{equation}
	\langle q_t^{ij}q_t^{kl}\rangle_P=\frac{1}{N(N-1)(N-2)(N-3)}\sum_{p_1}\sum_{p_2\ne p_1}\sum_{p_1'\ne p_1,p_2}\sum_{p_2'\ne p_1,p_2,p_1'}q_t^{p_1p_1'}q_t^{p_2p_2'},
\end{equation}
\begin{equation}
	\langle q_t^{ijkl}\rangle_P=\frac{1}{N(N-1)(N-2)(N-3)}\sum_{p_1}\sum_{p_2\ne p_1}\sum_{p_1'\ne p_1,p_2}\sum_{p_2'\ne p_1,p_2,p_1'}q_t^{p_1p_1'p_2p_2'}.
\end{equation}

We emphasize that, in our notation, whenever a similarity (or second order overlap) is between angular brackets (i.e., averaged), different indexes are indeed different. With all  these expressions in hand, we can write
\begin{align}
	&\mathbb{E}((q_{t+1}^{\alpha\beta})^2)\nonumber\\
	&=\frac{1}{B}-\frac{e^{-8\mu}}{4B}\left[\left(1+\frac{2}{N(N-1)}\right)+\left(2+\frac{4(N-2)}{N(N-1)}\right)\langle q_t^{p_1p'_1}\rangle_P+\frac{(N-2)(N-3)}{N(N-1)}\langle q_t^{p_1p_2p'_1p'_2}\rangle_P\right]\nonumber\\
	&+\frac{e^{-8\mu}}{4N}\left[\frac{N}{(N-1)}+\frac{(6N-8)}{(N-1)}\langle q_t^{p_1p'_1}\rangle_P+\left(N-1+\frac{1}{N-1}\right)\langle (q_t^{p_1p'_1})^2\rangle_P\right.\nonumber\\
	&+\left.\frac{2N(N-2)}{(N-1)}\langle q_t^{p_1p'_1}q_t^{p_1p'_2}\rangle_P+\frac{(N-2)(N-3)}{(N-1)}\langle q_t^{p_1p'_1}q_t^{p_2p'_2}\rangle_P\right].
\end{align}

In order to finish the calculation for the variance, we should subtract $\mathbb{E}[q_{t+1}^{\alpha\beta}]^2$. From equation \eqref{eq:reqmediasemqmin},
\begin{align}
	\mathbb{E}(q_{t+1}^{\alpha\beta})^2&=\left(\frac{e^{-4\mu}}{N^2}\sum_{p_1,p_1'}q_t^{p_1p_1'}\right)^2=\frac{e^{-8\mu}}{N^4}\sum_{p_1,p_1'}\sum_{p_2,p_2'}q_t^{p_1p_1'}q_t^{p_2p_2'}
\end{align}
and using the expansion of Eq.\eqref{eq:soma4},
\begin{align}
	\mathbb{E}(q_{t+1}^{\alpha\beta})^2&=\frac{e^{-8\mu}}{N^3}\left[N+2N(N-1)\langle q_t^{p_1p_1'}\rangle_P+2(N-1)\langle (q_t^{p_1p_1'})^2\rangle_P\right.\nonumber\\
	&\left.+4(N-1)(N-2)\langle q_t^{p_1p_1'}q_t^{p_1p_2'}\rangle_P+(N-1)(N-2)(N-3)\langle q_t^{p_1p_1'}q_t^{p_2p_2'}\rangle_P\right].
	\label{eq:mediaquadrado}
\end{align}
Hence, for the variance, we find
\begin{align}
	&\textrm{Var}(q_{t+1}^{\alpha\beta})\nonumber\\
	&=\frac{1}{B}-\frac{e^{-8\mu}}{4B}\left[\left(1+\frac{2}{N(N-1)}\right)+\left(2+\frac{4(N-2)}{N(N-1)}\right)\langle q_t^{p_1p'_1}\rangle_P+\frac{(N-2)(N-3)}{N(N-1)}\langle q_t^{p_1p_2p'_1p'_2}\rangle_P\right]\nonumber\\
	&+\frac{e^{-8\mu}(N-2)^2}{4N^2(N-1)}\left[1-2\langle q_t^{p_1p_1'}\rangle_P+\left(N+2-\frac{2}{N}\right)\langle (q_t^{p_1p_1'})^2\rangle_P\right.\nonumber\\
	&+\left.2\left(N-6+\frac{4}{N}\right)\langle q_t^{p_1p_1'}q_t^{p_1p_2'}\rangle_P-(N-3)\left(3-\frac{2}{N}\right)\langle q_t^{p_1p_1'}q_t^{p_2p_2'}\rangle_P\right].
	\label{eq:varParte1}
\end{align}

Now, considering the ensemble average of the equation above and the variance and covariance definitions
\begin{equation}
	\textrm{Var}(q_t^{\alpha\beta})=\left(\langle (q_t^{\alpha\beta})^2\rangle-\langle q_t^{\alpha\beta}\rangle^2\right)\Rightarrow\langle (q_t^{\alpha\beta})^2\rangle=\textrm{Var}(q_t^{\alpha\beta})+\langle q_t^{\alpha\beta}\rangle^2,
\end{equation}
\begin{equation}
	\textrm{Cov}(t)^{\alpha\beta\gamma}=\left(\langle q_t^{\alpha\beta}q_t^{\alpha\gamma}\rangle-\langle q_t^{\alpha\beta}\rangle^2\right)\Rightarrow\langle q_t^{\alpha\beta}q_t^{\alpha\gamma}\rangle=\textrm{Cov}(t)^{\alpha\beta\gamma}+\langle q_t^{\alpha\beta}\rangle^2,
\end{equation}
\begin{equation}
	\textrm{Cov}(t)^{\alpha\beta\gamma\delta}=\left(\langle q_t^{\alpha\beta}q_t^{\gamma\delta}\rangle-\langle q_t^{\alpha\beta}\rangle^2\right)\Rightarrow\langle q_t^{\alpha\beta}q_t^{\gamma\delta}\rangle=\textrm{Cov}(t)^{\alpha\beta\gamma\delta}+\langle q_t^{\alpha\beta}\rangle^2,
\end{equation}
and finally,
\begin{align}
	&\textrm{Var}(q_{t+1}^{\alpha\beta})\nonumber\\
	&=\frac{1}{B}-\frac{e^{-8\mu}}{4B}\left[\left(1+\frac{2}{N(N-1)}\right)+\left(2+\frac{4(N-2)}{N(N-1)}\right)\langle q_t^{p_1p'_1}\rangle+\frac{(N-2)(N-3)}{N(N-1)}\langle q_t^{p_1p_2p'_1p'_2}\rangle\right]\nonumber\\
	&+\frac{e^{-8\mu}(N-2)^2}{4N^2(N-1)}\left[\left(1-\langle q_t^{p_1p_1'}\rangle\right)^2+\left(N+2-\frac{2}{N}\right)\textrm{Var}(q_t^{\alpha\beta})\right.\nonumber\\
	&+\left.2\left(N-6+\frac{4}{N}\right)\textrm{Cov}(t)^{\alpha\beta\gamma}-(N-3)\left(3-\frac{2}{N}\right)\textrm{Cov}(t)^{\alpha\beta\gamma\delta}\right].
	\label{eq:varParte2}
\end{align}

Which is the recurrence equation (27) of the main text, in which we used $\textrm{Cov}(t)^{\alpha\beta\gamma\delta}=0$ (calculated ahead).

\section{The Second Order Overlap}

When calculating the variance of the similarity distribution, we defined the second order overlap $q^{\alpha\beta\gamma\delta}$ between the individuals $\alpha$, $\beta$, $\gamma$ and $\delta$,
\begin{equation}
	q_t^{\alpha\beta\gamma\delta}=\frac{1}{B}\sum_{i=1}^Bs_{i,t}^{\alpha}s_{i,t}^{\beta}s_{i,t}^{\gamma}s_{i,t}^{\delta}.
\end{equation}
We aim now to study its properties and evolution.

First, it is easy to see that when two individuals are the same, the second order overlap equals the \emph{first order overlap} (i.e., the similarity) between the remaining two individuals,
\begin{equation}
	q^{\alpha\alpha\gamma\delta}=\frac{1}{B}\sum_{i=1}^{B}s_i^{\alpha}s_i^{\alpha}s_i^{\gamma}s_i^{\delta}=\frac{1}{B}\sum_{i=1}^{B}s_i^{\gamma}s_i^{\delta}=q^{\gamma\delta}.
\end{equation}
Also, when there are two pairs of common individuals, the second overlap equals 1,
\begin{equation}
	q^{\alpha\alpha\gamma\gamma}=\frac{1}{B}\sum_{i=1}^{B}s_i^{\alpha}s_i^{\alpha}s_i^{\gamma}s_i^{\gamma}=\frac{1}{B}\sum_{i=1}^{B}1=1.
\end{equation}

These properties tell us that the interesting case to be considered is when the 4 individuals are different. Then, when writing any average of $q_t^{\alpha\beta\gamma\delta}$ we are gonna be referring to this case of interest.

To calculate the expected value $\mathbb{E}(q_{t+1}^{\alpha\beta\gamma\delta})$, we must know the distribution $\mathcal{P}(q_{t+1}^{\alpha\beta\gamma\delta})$, which is easily calculated once we remember that the genomes of different individuals are independent once we know their parents, as we have done for the mean similarity. Thus, the distribution is completely analogous to the one we have found for $q_{t+1}^{\alpha\beta}$, and it is given by
\begin{align}
	\mathcal{P}(q_{t+1}^{\alpha\beta\gamma\delta})&=\frac{1}{N^4}\sum_{p_1,p_2}\sum_{p_1',p_2'}\sum_{p_1'',p_2''}\sum_{p_1''',p_2'''}\frac{A_{p_1p_2}A_{p_1'p_2'}A_{p_1''p_2''}A_{p_1'''p_2'''}}{N_{p_1}N_{p_1'}N_{p_1''}N_{p_1'''}}\nonumber\\
	&\times\sum_{\mathbf{S}^{\alpha},\mathbf{S}^{\beta},\mathbf{S}^{\gamma},\mathbf{S}^{\delta}}\delta\left(q_{t+1}^{\alpha\beta\gamma\delta},\frac{1}{B}\sum_{j=1}^{B}s_j^{\alpha}s_j^{\beta}s_j^{\gamma}s_j^{\delta}\right)\nonumber\\
	&\times\prod_{i=1}^BF_{i,t}(\alpha,p_1,p_2)F_{i,t}(\beta,p_1',p_2')F_{i,t}(\gamma,p_1'',p_2'')F_{i,t}(\delta,p_1''',p_2'''),
\end{align}
where we have introduced the parents $(p_1'',p_2'')$ and $(p_1''',p_2''')$ of $\gamma$ and $\delta$, respectively.
Since all individuals are different, the expected value is obtained as
\begin{align}
	&\mathbb{E}(q_{t+1}^{\alpha\beta\gamma\delta})\nonumber\\
	&=\sum_{q}q\mathcal{P}(q_{t+1}^{\alpha\beta\gamma\delta}=q)\nonumber\\
	&=\frac{e^{-8\mu}}{16N^4}\sum_{p_1,p_2}\sum_{p_1',p_2'}\sum_{p_1'',p_2''}\sum_{p_1''',p_2'''}\frac{A_{p_1p_2}A_{p_1'p_2'}A_{p_1''p_2''}A_{p_1'''p_2'''}}{N_{p_1}N_{p_1'}N_{p_1''}N_{p_1'''}}\nonumber\\
	&\times\frac{1}{B}\sum_{j=1}^B(s_j^{p_1}+s_j^{p_2})(s_j^{p_1'}+s_j^{p_2'})(s_j^{p_1''}+s_j^{p_2''})(s_j^{p_1'''}+s_j^{p_2'''})\nonumber\\
	&=\frac{e^{-8\mu}}{16N^4}\sum_{p_1,p_2}\sum_{p_1',p_2'}\sum_{p_1'',p_2''}\sum_{p_1''',p_2'''}\frac{A_{p_1p_2}A_{p_1'p_2'}A_{p_1''p_2''}A_{p_1'''p_2'''}}{N_{p_1}N_{p_1'}N_{p_1''}N_{p_1'''}}\nonumber\\
	&\times
	\left(q_t^{p_1p_1'p_1''p_1'''}+q_t^{p_1p_1'p_1''p_2'''}+q_t^{p_1p_1'p_2''p_1'''}+q_t^{p_1p_1'p_2''p_2'''}\right.+q_t^{p_1p_2'p_1''p_1'''}+q_t^{p_1p_2'p_1''p_2'''}+q_t^{p_1p_2'p_2''p_1'''}+q_t^{p_1p_2'p_2''p_2'''}\nonumber\\
	&+q_t^{p_2p_1'p_1''p_1'''}+q_t^{p_2p_1'p_1''p_2'''}+q_t^{p_2p_1'p_2''p_1'''}+q_t^{p_2p_1'p_2''p_2'''}+\left.q_t^{p_2p_2'p_1''p_1'''}+q_t^{p_2p_2'p_1''p_2'''}+q_t^{p_2p_1'p_2''p_2'''}+q_t^{p_2p_2'p_2''p_2'''}\right),
\end{align}
which in the absence of assortative reproduction reduces to
\begin{equation}
	\mathbb{E}(q_{t+1}^{\alpha\beta\gamma\delta})=\frac{e^{-8\mu}}{N^4}\sum_{p_1,p_1',p_1'',p_1'''}q_t^{p_1p_1'p_1''p_1'''}\label{eq:meansecond}.
\end{equation}

Again, to deal with the four indexes sum, we use Eq.\eqref{eq:soma4} and find
\begin{align}
	\mathbb{E}(q_{t+1}^{\alpha\beta\gamma\delta})&=\frac{e^{-8\mu}}{N^3}\left[(3N-2)+(N-1)(6N-8)\langle q_t^{p_1p_1'}\rangle_P+(N-1)(N-2)(N-3)\langle q_t^{p_1p_1'p_1''p_1'''}\rangle_P\right]\\
	&\approx e^{-8\mu}\left[\frac{6}{N}\langle q_t^{p_1p_1'}\rangle_P+\left(1-\frac{6}{N}\right)\langle q_t^{p_1p_1'p_1''p_1'''}\rangle_P\right]
\end{align}
in which the approximation holds for $N\gg1$. The factor $6/N$ is the probability that two of the parents $p_1$, $p_1'$, $p_1''$ and $p_1'''$ are actually the same individual. Taking the ensemble average of this equation leads to the recurrence equation (26) of the main text.

\section{The similarity covariance}

The covariance appeared when we were calculating the expected value of $(q_{t+1}^{\alpha\beta})^2$. This is a very important piece of calculation. When changing from the genetic description of the individuals to the similarity description of the population, we are neglecting individual genome values and considering only a measure of their pairwise distance. Although it is a very natural change of description, since the dynamical constraint is given in terms of the genetic similarity, the probability of genomes of different individuals are independent of each other, while genetic similarities between different pairs of individuals may not be.

So let us start by defining the covariance between 
$q_t^{\alpha\beta}$ and $q_t^{\gamma\delta}$, with $\alpha\ne\beta$ and $\gamma\ne\delta$,
\begin{align}
	\textrm{Cov}(q_t^{\alpha\beta},q_t^{\gamma\delta})
	&=\mathbb{E}\left[(q_t^{\alpha\beta}-\mathbb{E}(q_t^{\alpha\beta}))(q_t^{\gamma\delta}-\mathbb{E}(q_t^{\gamma\delta}))\right]\nonumber\\
	&=\mathbb{E}(q_t^{\alpha\beta}q_t^{\gamma\delta})-\mathbb{E}(q_t^{\alpha\beta})\mathbb{E}(q_t^{\gamma\delta})\nonumber\\
	&=\mathbb{E}(q_t^{\alpha\beta}q_t^{\gamma\delta})-\mathbb{E}(q_t^{\alpha\beta})^2.
\end{align}

When both pairs of individuals are the same, $(\alpha,\beta)=(\gamma,\delta)$, the covariance equals the variance of $q^{\alpha\beta}_t$,
\begin{equation}
	\textrm{Cov}(q_t^{\alpha\beta},q_t^{\alpha\beta})=\textrm{Cov}(q_t^{\alpha\beta},q_t^{\beta\alpha})=\textrm{Var}(q_t^{\alpha\beta})
\end{equation}
and we have already performed this calculation. So now we are going to consider cases in which with all individuals are distinct or at most one of them is the same,
\begin{equation}
	\textrm{Cov}(q_t^{\alpha\beta},q_t^{\gamma\delta})\equiv\textrm{Cov}(t)^{\alpha\beta\gamma\delta}
\end{equation}
\begin{equation}
	\textrm{Cov}(q_t^{\alpha\beta},q_t^{\gamma\beta})\equiv\textrm{Cov}(t)^{\alpha\beta\gamma}
\end{equation}
in which $\alpha$, $\beta$, $\gamma$ and $\delta$ are all different.

\subsection{Covariance with one individual in common}

Let us start with the calculation of $\mathbb{E}(q_{t+1}^{\alpha\beta}q_{t+1}^{\gamma\beta})=\sum_{q_{t+1}^{\alpha\beta},q_{t+1}^{\gamma\beta}}q_{t+1}^{\alpha\beta}q_{t+1}^{\gamma\beta}\mathcal{P}(q_{t+1}^{\alpha\beta},q_{t+1}^{\gamma\beta})$, which introduces the task of calculating the joint distribution $\mathcal{P}(q_{t+1}^{\alpha\beta},q_{t+1}^{\gamma\beta})$. Following the same procedure we developed in the main text, since we learned that the genomes from different individuals can be treated as independent when conditioned to the individuals' parents, it is possible to write such distribution as
\begin{align}
	\mathcal{P}(q_{t+1}^{\alpha\beta},q_{t+1}^{\gamma\beta})&=\frac{1}{N^3}\sum_{p_1,p_2}\sum_{p_1',p_2'}\sum_{p_1'',p_2''}\frac{A_{p_1p_2}A_{p_1'p_2'}A_{p_1''p_2''}}{N_{p_1}N_{p_1'}N_{p_1''}}\sum_{\mathbf{S}^{\alpha},\mathbf{S}^{\beta},\mathbf{S}^{\gamma}}\delta\left(q_{t+1}^{\alpha\beta},\frac{\mathbf{S}^{\alpha}\cdot\mathbf{S}^{\beta}}{B}\right)\delta\left(q_{t+1}^{\gamma\beta},\frac{\mathbf{S}^{\gamma}\cdot\mathbf{S}^{\beta}}{B}\right)\nonumber\\
	&\times\prod_{i=1}^BF_{i,t}(\alpha,p_1,p_2)F_{i,t}(\beta,p_1',p_2')F_{i,t}(\gamma,p_1'',p_2''),
\end{align}
where we introduced the parents $(p_1'',p_2'')$ of the individual $\gamma$. Then,
\begin{align}
	\mathbb{E}(q_{t+1}^{\alpha\beta}q_{t+1}^{\gamma\beta})&= \sum_{q_{t+1}^{\alpha\beta},q_{t+1}^{\gamma\beta}}q_{t+1}^{\alpha\beta}q_{t+1}^{\gamma\beta}\mathcal{P}(q_{t+1}^{\alpha\beta},q_{t+1}^{\gamma\beta})\nonumber\\
	&=\frac{1}{N^3}\sum_{p_1,p_2}\sum_{p_1',p_2'}\sum_{p_1'',p_2''}\frac{A_{p_1p_2}A_{p'_1p'_2}A_{p''_1p''_2}}{N_{p_1}N_{p'_1}N_{p''_1}}\sum_{\mathbf{S}^{\alpha},\mathbf{S}^{\beta},\mathbf{S}^{\gamma}}\left(\frac{\mathbf{S}^{\alpha}\cdot\mathbf{S}^{\beta}}{B}\right)\left(\frac{\mathbf{S}^{\gamma}\cdot\mathbf{S}^{\beta}}{B}\right)\nonumber\\
	&\times\prod_{i=1}^BF_{i,t}(\alpha,p_1,p_2)F_{i,t}(\beta,p'_1,p'_2)F_{i,t}(\gamma,p''_1,p''_2)\nonumber\\
	&=\frac{1}{N^3B^2}\sum_{p_1,p_2}\sum_{p'_1,p'_2}\sum_{p''_1,p''_2}\frac{A_{p_1p_2}A_{p'_1p'_2}A_{p''_1p''_2}}{N_{p_1}N_{p'_1}N_{p''_1}}\sum_{\mathbf{S}^{\alpha},\mathbf{S}^{\beta},\mathbf{S}^{\gamma}}\left(\sum_{j,k}s_j^{\alpha}s_j^{\beta}s_k^{\gamma}s_k^{\beta}\right)\nonumber\\
	&\times\prod_{i=1}^BF_{i,t}(\alpha,p_1,p_2)F_{i,t}(\beta,p'_1,p'_2)F_{i,t}(\gamma,p''_1,p''_2)\nonumber\\
	&=\frac{1}{N^3B^2}\sum_{p_1,p_2}\sum_{p'_1,p'_2}\sum_{p''_1,p''_2}\frac{A_{p_1p_2}A_{p'_1p'_2}A_{p''_1p''_2}}{N_{p_1}N_{p'_1}N_{p''_1}}\nonumber\\
	&\times\sum_{\mathbf{S}^{\alpha},\mathbf{S}^{\beta},\mathbf{S}^{\gamma}}\sum_{j=1}^B\left(s_j^{\alpha}s_j^{\gamma}+\sum_{k\ne j}s_j^{\alpha}s_j^{\beta}s_k^{\gamma}s_k^{\beta}\right)\prod_{i=1}^BF_{i,t}(\alpha)F_{i,t}(\beta)F_{i,t}(\gamma).
\end{align}
The last line of this equation is calculated with the same technique used before,
\begin{align}
	&\sum_{\mathbf{S}^{\alpha},\mathbf{S}^{\beta},\mathbf{S}^{\gamma}}\sum_{j=1}^B\left(s_j^{\alpha}s_j^{\gamma}+\sum_{k\ne j}s_j^{\alpha}s_j^{\beta}s_k^{\gamma}s_k^{\beta}\right)\prod_{i=1}^BF_{i,t}(\alpha)F_{i,t}(\beta)F_{i,t}(\gamma)\nonumber\\
	&=\sum_{j=1}^B\frac{e^{-4\mu}}{4}(s_{j}^{p_1}+s_{j}^{p_2})(s_{j}^{p_1''}+s_{j}^{p_2''})+\sum_{j=1}^B\sum_{k\ne j}\frac{e^{-8\mu}}{16}(s_{j}^{p_1}+s_{j}^{p_2})(s_{j}^{p_1'}+s_{j}^{p_2'})(s_{k}^{p_1''}+s_{k}^{p_2''})(s_{k}^{p_1'}+s_{k}^{p_2'})\nonumber\\
	&=\sum_{j=1}^B\frac{e^{-4\mu}}{4}(s_{j}^{p_1}+s_{j}^{p_2})(s_{j}^{p_1''}+s_{j}^{p_2''})+\sum_{j=1}^B\frac{e^{-8\mu}}{16}(s_{j}^{p_1}+s_{j}^{p_2})(s_{j}^{p_1'}+s_{j}^{p_2'})\sum_{k=1}^B(s_{k}^{p_1''}+s_{k}^{p_2''})(s_{k}^{p_1'}+s_{k}^{p_2'})\nonumber\\
	&-\sum_{j=1}^B\frac{e^{-8\mu}}{16}(s_{j}^{p_1}+s_{j}^{p_2})(s_{j}^{p_1'}+s_{j}^{p_2'})(s_{j}^{p_1''}+s_{j}^{p_2''})(s_{j}^{p_1'}+s_{j}^{p_2'})\nonumber\\
	&=B\frac{e^{-4\mu}}{4}(q_t^{p_1p_1''}+q_t^{p_1p_2''}+q_t^{p_2p_1''}+q_t^{p_2p_2''})\nonumber\\
	&+B^2\frac{e^{-8\mu}}{16}(q_t^{p_1p_1'}+q_t^{p_1p_2'}+q_t^{p_2p_1'}+q_t^{p_2p_2'})(q_t^{p_1''p_1'}+q_t^{p_1''p_2'}+q_t^{p_2''p_1'}+q_t^{p_2''p_2'})\nonumber\\
	&-B\frac{e^{-8\mu}}{8}\left[(q_t^{p_1p_1''}+q_t^{p_1p_2''}+q_t^{p_2p_1''}+q_t^{p_2p_2''})+(q_t^{p_1p_1''p_1'p_2'}+q_t^{p_1p_2''p_1'p_2'}+q_t^{p_2p_1''p_1'p_2'}+q_t^{p_2p_2''p_1'p_2'})\right].
\end{align}

Now, we can write
\begin{align}
	\mathbb{E}(q_{t+1}^{\alpha\beta}q_{t+1}^{\gamma\beta})&=\frac{1}{N^3}\sum_{p_1,p_2}\sum_{p'_1,p'_2}\sum_{p''_1,p''_2}\frac{A_{p_1p_2}A_{p'_1p'_2}A_{p''_1p''_2}}{N_{p_1}N_{p'_1}N_{p''_1}}\nonumber\\
	&\times\left\{\frac{e^{-8\mu}}{16}(q_t^{p_1p_1'}+q_t^{p_1p_2'}+q_t^{p_2p_1'}+q_t^{p_2p_2'})(q_t^{p_1''p_1'}+q_t^{p_1''p_2'}+q_t^{p_2''p_1'}+q_t^{p_2''p_2'})\right.\nonumber\\
	&+\frac{1}{B}\left[\frac{e^{-4\mu}}{4}-\frac{e^{-8\mu}}{8}\right](q_t^{p_1p_1''}+q_t^{p_1p_2''}+q_t^{p_2p_1''}+q_t^{p_2p_2''})\nonumber\\
	&\left.-\frac{e^{-8\mu}}{8B}(q_t^{p_1p_1''p_1'p_2'}+q_t^{p_1p_2''p_1'p_2'}+q_t^{p_2p_1''p_1'p_2'}+q_t^{p_2p_2''p_1'p_2'})\right\}.\label{eq:secondmoment2}
\end{align}

\subsection{Covariance with one individual in common and no assortative reproduction}

As we did for the variance, we shall now consider the case without assortative reproduction. Although in this case there is a sum over six indexes, a maximum of only four indexes appear in each term,
\begin{align}
	\mathbb{E}(q_{t+1}^{\alpha\beta}q_{t+1}^{\gamma\beta})&=\frac{1}{N^3(N-1)^3}\sum_{p_1,p_2}\sum_{p'_1,p'_2}\sum_{p''_1,p''_2}A_{p_1p_2}A_{p'_1p'_2}A_{p''_1p''_2}\nonumber\\
	&\times\left\{\frac{e^{-8\mu}}{2}(q_t^{p_1p_1'}q_t^{p_1'p_1''}+q_t^{p_1p_1'}q_t^{p_2'p_2''})\right.+\frac{1}{B}\left[e^{-4\mu}-\frac{e^{-8\mu}}{2}\right]q_t^{p_1p_1''}\left.-\frac{e^{-8\mu}}{2B}q_t^{p_1p_1''p_1'p_2'}\right\}\nonumber\\
	&=\frac{1}{B}\left[e^{-4\mu}-\frac{e^{-8\mu}}{2}\right]\left[\frac{1}{N}+\left(1-\frac{1}{N}\right)\langle q_t^{p_1p_1''}\rangle_P\right]\nonumber\\
	&+\frac{e^{-8\mu}}{2N^3}\sum_{p_1,p'_1,p''_1}q_t^{p_1p_1'}q_t^{p_1'p_1''}+\frac{e^{-8\mu}}{2N^3(N-1)}\sum_{p_1,p'_1,p'_2,p''_1}A_{p'_1p'_2}q_t^{p_1p_1'}q_t^{p_2'p_1''}\nonumber\\
	&-\frac{e^{-8\mu}}{2BN^3(N-1)}\sum_{p_1,p_1',p_2',p_1''}A_{p_1'p_2'}q_t^{p_1p_1''p_1'p_2'},
\end{align}
and using the expansions in equations \eqref{eq:soma3} and \eqref{eq:soma4}, we find, for each one of the sums,
\begin{align}
	\sum_{p_1,p'_1,p''_1}q_t^{p_1p_1'}q_t^{p_1'p_1''}&=N(N-1)\left[\frac{1}{N-1}+2\langle q_t^{p_1p_1'}\rangle_P+\langle(q_t^{p_1p_1'})^2\rangle_P+(N-2)\langle q_t^{p_1p_1'}q_t^{p_1'p_1''}\rangle_P\right],
\end{align}
\begin{align}
	\sum_{p_1,p'_1,p'_2,p''_1}A_{p'_1p'_2}q_t^{p_1p_1'}q_t^{p_2'p_1''}&=N(N-1)\left[1+2(N-1)\langle q_t^{p_1p_1'}\rangle_P+\langle (q_t^{p_1p_1'})^2\rangle_P\right.\nonumber\\
	&+\left.(N-2)(N-3)\langle q_t^{p_1p_1'}q_t^{p_2p_2'}\rangle_P+3(N-2)\langle q_t^{p_1p_1'}q_t^{p_1'p_1''}\rangle_P\right],
\end{align}
\begin{align}
	\sum_{p_1,p'_1,p'_2,p''_1}A_{p'_1p'_2}q_t^{p_1p_1''p_1'p_2'}&=N(N-1)\left[2+(5N-8)\langle q_t^{p_1p_1'}\rangle_P+(N-2)(N-3)\langle q_t^{p_1p_2p_1'p_2'}\rangle_P\right].
\end{align}
Combining all these results and subtracting $\mathbb{E}(q_{t+1}^{\alpha\beta})^2$ (given by Eq.\eqref{eq:mediaquadrado}) we find
\begin{align}
	\textrm{Cov}(t+1)^{\alpha\beta\gamma}
	&=\frac{e^{-4\mu}}{B}\left[\frac{1}{N}+\left(1-\frac{1}{N}\right)\langle q_t^{p_1p_1'}\rangle_P\right]\nonumber\\
	&-\frac{e^{-8\mu}}{2B}\left[\frac{2}{N^2}+\frac{1}{N}+\left(1+\frac{4}{N}-\frac{8}{N^2}\right)\langle q_t^{p_1p_1'}\rangle_P+\left(1-\frac{2}{N}\right)\left(1-\frac{3}{N}\right)\langle q_t^{p_1p_2p_1'p_2'}\rangle_P\right]\nonumber\\
	&+\frac{e^{-8\mu}(n-2)^2}{2N^3}\left[\langle(q_t^{p_1p_1'})^2\rangle_P+(N-4)\langle q_t^{p_1p_1'}q_t^{p_1p_2'}\rangle_P^2-(N-3)\langle q_t^{p_1p_1'}q_t^{p_2p_2'}\rangle_P\right]\nonumber\\
	&=\frac{e^{-4\mu}}{B}\left[\frac{1}{N}+\left(1-\frac{1}{N}\right)\langle q_t^{p_1p_1'}\rangle_P\right]\nonumber\\
	&-\frac{e^{-8\mu}}{2B}\left[\frac{2}{N^2}+\frac{1}{N}+\left(1+\frac{4}{N}-\frac{8}{N^2}\right)\langle q_t^{p_1p_1'}\rangle_P+\left(1-\frac{2}{N}\right)\left(1-\frac{3}{N}\right)\langle q_t^{p_1p_2p_1'p_2'}\rangle_P\right]\nonumber\\
	&+\frac{e^{-8\mu}(n-2)^2}{2N^3}\left(1-\frac{1}{\overline{N}}\right)\left[\textrm{Var}(q_t^{p_1p_1'})_P+(N-4)\textrm{Cov}(t)^{\alpha\beta\gamma}_P-(N-3)\textrm{Cov}(t)^{\alpha\beta\gamma\delta}_P\right].\label{eq:covariancia}
\end{align}
Taking the ensemble average of this equation, we finally find the result of equation (29) from the main text.

\subsection{Covariance with no individual in common}

To calculate this quantity requires the joint distribution $\mathcal{P}(q_{t+1}^{\alpha\beta}q_{t+1}^{\gamma\delta})$, which can be calculated analogously as before,
\begin{align}
	\mathcal{P}(q_{t+1}^{\alpha\beta},q_{t+1}^{\gamma\delta})&=\frac{1}{N^4}\sum_{p_1,p_2}\sum_{p_1',p_2'}\sum_{p_1'',p_2''}\sum_{p_1''',p_2'''}\frac{A_{p_1p_2}A_{p_1'p_2'}A_{p_1''p_2''}A_{p_1'''p_2'''}}{N_{p_1}N_{p_1'}N_{p_1''}N_{p_1'''}}\nonumber\\
	&\times\sum_{\mathbf{S}^{\alpha},\mathbf{S}^{\beta},\mathbf{S}^{\gamma},\mathbf{S}^{\delta}}\delta\left(q_{t+1}^{\alpha\beta},\frac{\mathbf{S}^{\alpha}\cdot\mathbf{S}^{\beta}}{B}\right)\delta\left(q_{t+1}^{\gamma\delta},\frac{\mathbf{S}^{\gamma}\cdot\mathbf{S}^{\delta}}{B}\right)\nonumber\\
	&\times\prod_{i=1}^BF_{i,t}(\alpha,p_1,p_2)F_{i,t}(\beta,p_1',p_2')F_{i,t}(\gamma,p_1'',p_2'')F_{i,t}(\delta,p_1''',p_2'''),
\end{align}
where we introduced the parents $(p_1''',p_2''')$ of the individual	$\delta$. Then,
\begin{align}
	\mathbb{E}(q_{t+1}^{\alpha\beta}q_{t+1}^{\gamma\delta})&= \sum_{q_{t+1}^{\alpha\beta},q_{t+1}^{\gamma\delta}}q_{t+1}^{\alpha\beta}q_{t+1}^{\gamma\delta}\mathcal{P}(q_{t+1}^{\alpha\beta},q_{t+1}^{\gamma\beta})\nonumber\\
	&=\frac{1}{N^4}\sum_{p_1,p_2}\sum_{p_1',p_2'}\sum_{p_1'',p_2''}\sum_{p_1''',p_2'''}\frac{A_{p_1p_2}A_{p'_1p'_2}A_{p''_1p''_2}A_{p'''_1p'''_2}}{N_{p_1}N_{p'_1}N_{p''_1}N_{p'''_1}}\sum_{\mathbf{S}^{\alpha},\mathbf{S}^{\beta},\mathbf{S}^{\gamma},\mathbf{S}^{\delta}}\left(\frac{\mathbf{S}^{\alpha}\cdot\mathbf{S}^{\beta}}{B}\right)\left(\frac{\mathbf{S}^{\gamma}\cdot\mathbf{S}^{\delta}}{B}\right)\nonumber\\
	&\times\prod_{i=1}^BF_{i,t}(\alpha,p_1,p_2)F_{i,t}(\beta,p'_1,p'_2)F_{i,t}(\gamma,p''_1,p''_2)F_{i,t}(\delta,p'''_1,p'''_2)\nonumber\\
	&=\frac{1}{N^2}\sum_{p_1,p_2}\sum_{p_1',p_2'}\frac{A_{p_1p_2}A_{p'_1p'_2}}{N_{p_1}N_{p'_1}}\sum_{\mathbf{S}^{\alpha},\mathbf{S}^{\beta}}\left(\frac{\mathbf{S}^{\alpha}\cdot\mathbf{S}^{\beta}}{B}\right)\prod_{i=1}^BF_{i,t}(\alpha,p_1,p_2)F_{i,t}(\beta,p'_1,p'_2)\nonumber\\
	&\times\frac{1}{N^2}\sum_{p_1'',p_2''}\sum_{p_1''',p_2'''}\frac{A_{p_1''p_2''}A_{p'''_1p'''_2}}{N_{p_1''}N_{p'''_1}}\sum_{\mathbf{S}^{\gamma},\mathbf{S}^{\delta}}\left(\frac{\mathbf{S}^{\gamma}\cdot\mathbf{S}^{\delta}}{B}\right)\prod_{i=1}^BF_{i,t}(\gamma,p_1'',p_2'')F_{i,t}(\delta,p'''_1,p'''_2)\nonumber\\
	&=\mathbb{E}(q_{t+1}^{\alpha\beta})\mathbb{E}(q_{t+1}^{\gamma\delta}).
\end{align}

This result shows that the covariance of similarities that do not share common individuals is zero,
\begin{equation}
	\textrm{Cov}(t)^{\alpha\beta\gamma\delta}=	\mathbb{E}(q_{t+1}^{\alpha\beta}q_{t+1}^{\gamma\delta})-\mathbb{E}(q_{t+1}^{\alpha\beta})\mathbb{E}(q_{t+1}^{\gamma\delta})=0.
\end{equation}

\section{The dependence on the genome size $B$}

Notice that the variance $\textrm{Var}(q_{t+1}^{\alpha\beta})$ in equation \eqref{eq:varParte2} has the form
\begin{equation}
	\textrm{Var}(q_{t+1}^{\alpha\beta})=a_t+c_{1}\textrm{Var}(q_{t}^{\alpha\beta})+c_{2}\textrm{Cov}(t)^{\alpha\beta\gamma}+\frac{b_t}{B}, \nonumber
\end{equation}
where $a_t$ and $b_t$ depend and $c_1$ and $c_2$ are constants, but none depend on $B$. The same happens for the covariance $\textrm{Cov}(t)^{\alpha\beta\gamma}$ in equation \eqref{eq:covariancia}.
\begin{equation}
	\textrm{Cov}(t+1)^{\alpha\beta\gamma}=c_{3}\textrm{Var}(q_{t}^{\alpha\beta})+c_{4}\textrm{Cov}(t)^{\alpha\beta\gamma}+\frac{b'_t}{B}, \nonumber
\end{equation}
where $c_3$ and $c_4$ are constants and $b_t'$ depend on time, but none on $B$.

Solving one step in time of this recurrence is enough to see that terms over $B$ accumulate:
\begin{align}
	\textrm{Var}(q_{t+1}^{\alpha\beta})=&a_t+c_{1}\left(a_{t-1}+c_{1}\textrm{Var}(q_{t-1}^{\alpha\beta})+c_{2}\textrm{Cov}(t-1)^{\alpha\beta\gamma}+\frac{b_{t-1}}{B}\right)\nonumber\\
	&+c_{2}\left(c_{3}\textrm{Var}(q_{t-1}^{\alpha\beta})+c_{4}\textrm{Cov}(t-1)^{\alpha\beta\gamma}+\frac{b'_{t-1}}{B}\right)+\frac{b_t}{B}\nonumber\\
	=&(a_t+c_1a_{t-1})+(c_1^2+c_2c_3)\textrm{Var}(q_{t-1}^{\alpha\beta})+(c_1c_2+c_2c_4)\textrm{Cov}(t-1)^{\alpha\beta\gamma}\nonumber\\
	&+\frac{(c_1b_{t-1}+c_2b'_{t-1}+b_t)}{B}
\end{align}
and analogous for $\textrm{Cov}(t)^{\alpha\beta\gamma}$. Therefore, the variance as a function of $B$ can always be expressed as
\begin{equation}
	\textrm{Var}(q_{t}^{\alpha\beta})=\Lambda_1(t)+\frac{\Lambda_2(t)}{B},
\end{equation}
where the functions of time $\Lambda_1(t)$ and $\Lambda_2(t)$ are found by recursively solving the expressions for the variance and the covariance (as also for the mean overlap and mean second order overlap). This last equation is the Eq. (19) in the main text.

\section{On higher-order overlaps}

The Derrida-Higgs model, as we are describing it, can be understood as an indexed family of random vectors $\boldsymbol{Q}_t$ defined as
\begin{equation}
	\boldsymbol{Q}_t\equiv\{q_t^{12},q_t^{13},\ldots,q_t^{1N},q_t^{23},q_t^{24},\ldots,q_t^{2N},\ldots,q_t^{N-1,N}\}=\{q_t^{ij}|1\le i<j\le N\},
\end{equation}
i.e., the upper triangle of the similarity matrix $\mathbb{Q}_t$, whose entries follow a distribution with mean and covariance which we have calculated in the previous section. In fact, the probability distributions we have calculated so far are marginals of the distribution $\mathcal{P}(\boldsymbol{Q}_{t+1})$. For instance,
\begin{equation}
	\mathcal{P}(q_{t+1}^{\alpha\beta})=\sum_{q_{t+1}^{12}}\cdots\sum_{q_{t+1}^{\alpha1}}\cdots\sum_{q_{t+1}^{\alpha,\beta-1}}\sum_{q_{t+1}^{\alpha,\beta+1}}\cdots\sum_{q_{t+1}^{N-1,N}}\mathcal{P}(\boldsymbol{Q}_{t+1}),
\end{equation}
with the property of being the same regardless of $\alpha$ and $\beta$ (with $\alpha\ne\beta$). Using the procedure we have introduced to calculate $\mathcal{P}(q_{t+1}^{\alpha\beta})$, we can carry on the calculation of $\mathcal{P}(\boldsymbol{Q}_{t+1})$. Let $p_1(\alpha_i)$ be the focal parent of the individual $\alpha_i$, and $p_2(\alpha_i)$ its second parent ($p_1$ and $p_2$ are taken from the generation $t$ while $\alpha_i$ is from generation $t+1$). Then $\mathcal{P}(\boldsymbol{Q}_{t+1})$ takes the form
\begin{align}
	\mathcal{P}(\boldsymbol{Q}_{t+1})&=\frac{1}{N^{N}}\sum_{p_1(\alpha_1),p_2(\alpha_1)}\cdots\sum_{p_1(\alpha_N),p_2(\alpha_N)}\left(\prod_{i=1}^{N}\frac{A_{p_1(\alpha_i)p_2(\alpha_i)}}{N_{p_1(\alpha_i)}}\right)\nonumber\\
	&\times\sum_{\mathbf{S}^1}\cdots\sum_{\mathbf{S}^N}\left[\prod_{\{q_{t+1}^{ij}|1\le i<j\le N\}}\delta\left(q_{t+1}^{ij},\frac{\mathbf{S}^i\cdot\mathbf{S}^j}{B}\right)\right]\left[\prod_{k=1}^B\left(\prod_{l=1}^NF_{k,t}(\alpha_l,p_1(\alpha_l),p_2(\alpha_l))\right)\right].
\end{align}

On the other hand, as we have seen, second moments of $\mathcal{P}(\boldsymbol{Q}_{t+1})$ also depend on the second order overlap, and it is not hard to see that a third moment of this distribution would also depend on the \emph{third order overlap}, and so on. For instance, let us consider the third moment
$\mathbb{E}(q_{t+1}^{\alpha\beta}q_{t+1}^{\gamma\beta}q_{t+1}^{\gamma\alpha})$, whose involved probability is
\begin{align}
	&\mathcal{P}(q_{t+1}^{\alpha\beta},q_{t+1}^{\gamma\beta},q_{t+1}^{\gamma\alpha})\nonumber\\
	&=\frac{1}{N^3}\sum_{p_1,p_2}\sum_{p_1',p_2'}\sum_{p_1'',p_2''}\frac{A_{p_1p_2}A_{p_1'p_2'}A_{p_1''p_2''}}{N_{p_1}N_{p_1'}N_{p_1''}}\sum_{\mathbf{S}^{\alpha},\mathbf{S}^{\beta},\mathbf{S}^{\gamma}}\delta\left(q_{t+1}^{\alpha\beta},\frac{\mathbf{S}^{\alpha}\cdot\mathbf{S}^{\beta}}{B}\right)\delta\left(q_{t+1}^{\gamma\beta},\frac{\mathbf{S}^{\gamma}\cdot\mathbf{S}^{\beta}}{B}\right)\delta\left(q_{t+1}^{\gamma\alpha},\frac{\mathbf{S}^{\gamma}\cdot\mathbf{S}^{\alpha}}{B}\right)\nonumber\\
	&\times\prod_{i=1}^BF_{i,t}(\alpha,p_1,p_2)F_{i,t}(\beta,p_1',p_2')F_{i,t}(\gamma,p_1'',p_2''),
\end{align}
and thus, when calculating it, we find the term
\begin{align}
	&\frac{1}{B^3}\sum_{\mathbf{S}^{\alpha},\mathbf{S}^{\beta},\mathbf{S}^{\gamma}}\sum_{j,k,l}s_j^{\alpha}s_j^{\beta}s_k^{\beta}s_k^{\gamma}s_l^{\gamma}s_l^{\alpha}\prod_{i=1}^BF_{i,t}(\alpha,p_1,p_2)F_{i,t}(\beta,p_1',p_2')F_{i,t}(\gamma,p_1'',p_2'')\nonumber\\
	=&\frac{1}{B^3}\left(\frac{e^{-2\mu}}{2}\right)^6\sum_{j,k,l}(s_j^{p_1}+s_j^{p_2})(s_j^{p_1'}+s_j^{p_2'})(s_k^{p_1'}+s_k^{p_2'})(s_k^{p_1''}+s_k^{p_2''})(s_l^{p_1''}+s_l^{p_2''})(s_l^{p_1}+s_l^{p_2})\nonumber\\
	=&\frac{1}{B^3}\left(\frac{e^{-2\mu}}{2}\right)^6\left[\ldots+8\sum_ks_k^{p_1}s_k^{p_2}s_k^{p_1'}s_k^{p_2'}s_k^{p_1''}s_k^{p_2''}\right],
\end{align}
where we can recognize the third order overlap,
\begin{equation}
	q^{p_1p_2p_1'p_2'p_1''p_2''}\equiv\frac{1}{B}\sum_ks_k^{p_1}s_k^{p_2}s_k^{p_1'}s_k^{p_2'}s_k^{p_1''}s_k^{p_2''}.
\end{equation}

Thus, it is not possible to completely change from the genome description to \emph{only} the first order overlap description, once the evolution of its distribution depends on higher order overlaps.

\subsection{The definition of a higher-order overlap}

For completeness, let us introduce a general definition for the overlap. Let $n$ individuals $\{\alpha_1,\alpha_2,\ldots,\alpha_n\}$ all with their own genome $\{s_1^{\alpha_k},\ldots,s_B^{\alpha_k}\}$. The $j-$order overlap of the $2j$ individuals $\{\alpha_{i_1},\ldots,\alpha_{i_{2j}}\}$ with $\{i_1,\ldots,i_{2j}\}\subset\{1,\ldots,n\}$ is defined by
\begin{equation}
	q^{(j)}(i_1,\ldots,i_{2j})\equiv\frac{1}{B}\sum_{k=1}^Bs_k^{\alpha_{i_1}}s_k^{\alpha_{i_2}}\ldots s_k^{\alpha_{i_{2j}}}.
\end{equation}
Notice that we have changed  the notation from what we have used so far, since carrying the individuals as upper indexes can be quite messy for higher order overlaps.

\subsection{Properties}

Let us now introduce some of its properties.

\subsubsection{Identity property}

If all individuals have the same genome, then the overlap (of any order) equals 1,
\begin{equation}
	q^{(j)}(i_1,\ldots,i_{2j})=\frac{1}{B}\sum_{k=1}^Bs_k^{\alpha_{i_1}}s_k^{\alpha_{i_2}}\ldots s_k^{\alpha_{i_{2j}}}=\frac{1}{B}\sum_{k=1}^B 1=1.
\end{equation}

\subsubsection{Permutation symmetry}

The overlap (of any order) does not change under permutations of individuals,
\begin{align}
	q^{(j)}(\ldots\alpha_{l},\ldots,\alpha_{m}\ldots)&=\frac{1}{B}\sum_{k=1}^B\ldots s_k^{\alpha_{l}}\ldots s_k^{\alpha_{m}}\ldots\nonumber\\
	&=\frac{1}{B}\sum_{k=1}^B\ldots s_k^{\alpha_{m}}\ldots s_k^{\alpha_{l}}\ldots\nonumber\\
	&=q^{(j)}(\ldots\alpha_{m},\ldots,\alpha_{l}\ldots).
\end{align}

\subsubsection{The reduced order property}

If two individuals of the set $\{\alpha_{i_1},\ldots,\alpha_{i_{2j}}\}$ are the same, then the $j-$order overlap equals the $(j-1)-$order overlap of the same set without these two individuals. Suppose $\alpha_l=\alpha_m$. Then,
\begin{align}
	q^{(j)}(\ldots,\alpha_{l-1},\alpha_{l},\alpha_{l+1},\ldots\,\alpha_{m-1},\alpha_{m},\alpha_{l+1}\ldots)&=\frac{1}{B}\sum_{k=1}^B\ldots s_k^{\alpha_{l-1}}s_k^{\alpha_{l}}s_k^{\alpha_{l+1}}\ldots s_k^{\alpha_{m-1}}s_k^{\alpha_{m}}s_k^{\alpha_{m+1}}\ldots\nonumber\\
	&=\frac{1}{B}\sum_{k=1}^B\left(\ldots s_k^{\alpha_{l-1}}s_k^{\alpha_{l+1}}\ldots s_k^{\alpha_{m-1}}s_k^{\alpha_{m+1}}\ldots\right)s_k^{\alpha_{l}}s_k^{\alpha_{m}}\nonumber\\
	&=\frac{1}{B}\sum_{k=1}^B\left(\ldots s_k^{\alpha_{l-1}}s_k^{\alpha_{l+1}}\ldots s_k^{\alpha_{m-1}}s_k^{\alpha_{m+1}}\ldots\right)\nonumber\\
	&=q^{(j-1)}(\ldots,\alpha_{l-1},\alpha_{l+1},\ldots\,\alpha_{m-1},\alpha_{l+1}\ldots).
\end{align}
Indeed, if $\{\alpha_{i_1},\ldots,\alpha_{i_{2j}}\}$ has $m$ pairs of equal individuals, then its $j-$order overlap equals the $(j-m)-$order of the same set excluding the $m$ pairs.

\subsubsection{First order mean evolution}

In the absence of $q_{min}$, we can approximate the evolution of the $j-$order overlap as follows. Considering $\{\alpha_1,\ldots,\alpha_{2j}\}$ of $2j$ different individuals in a population of size $N$, and extending the result of equation \eqref{eq:meansecond},
we can write
\begin{equation}
	\mathbb{E}(q^{(j)}(\alpha_1,\ldots,\alpha_{2j}))=e^{-4j\mu}\left(\frac{1}{N}\right)^{2j}\sum_{p_1(\alpha_1),\ldots,p_1(\alpha_{2j})}q^{(j)}(p_1(\alpha_1),\ldots,p_1(\alpha_{2j})).
\end{equation}
We can approximate the last two factors of this expression if we remember that two individuals of the set can be the same $\{(p_1(\alpha_1),\ldots,p_1(\alpha_{2j})\}$ and then use the reduced order property. If one is going to form a set of entries, there can be a pair of the same individual on $2j(2j-1)/2!$ different positions. Choosing the entries at random (individuals) from a total of $N$, given an individual, the chance  of choosing another one equal to the first is $1/N$. Then,
\begin{equation}
	\mathbb{E}(q_{t+1}^{(j)})\approx e^{-4j\mu}\left[\frac{j(2j-1)}{N}\langle q_t^{(j-1)}\rangle_P+\left(1-\frac{j(2j-1)}{N}\right)\langle q_t^{(j)}\rangle_P\right],
\end{equation}
which approximates the evolution up to order $1/N$.

\endgroup

\end{document}